\documentclass[12pt,epsf]{article}

\usepackage{times}
\usepackage{graphicx}
\usepackage{amsfonts}
\usepackage{amsmath}
\usepackage{amssymb}
\usepackage{amstext}
\usepackage{amsthm}
\usepackage{youngtab}

\usepackage{epsfig}
\usepackage{color}
\usepackage{leftidx}
\usepackage{tikz}
\psfigdriver{dvips}
\usepackage{latexsym}
\usepackage[matrix,curve]{xypic}
\usepackage{rotating}
\rotdriver{dvips}

\begin{document}

\baselineskip 6mm
\renewcommand{\thefootnote}{\fnsymbol{footnote}}

%------------ Hyun Seok's macro's, etc  -----------

\newcommand{\nc}{\newcommand}
\newcommand{\rnc}{\renewcommand}

%\headheight=0truein
%\headsep=0truein
%\topmargin=0truein
%\oddsidemargin=0truein
%\evensidemargin=0truein
%\textheight=9truein
%\textwidth=6.5truein

\rnc{\baselinestretch}{1.24}    % 1.5 spacing btwn text lines
\setlength{\jot}{6pt}       % spacing btwn the rows of an eqnarray
\rnc{\arraystretch}{1.24}   % spacing btwn the rows of a non-eqn array

%%%%%%%%%%%%%%%%%%%%%% Equation Numbering %%%%%%%%%%%%%%%%%%%%%%%
\makeatletter
\rnc{\theequation}{\thesection.\arabic{equation}}
\@addtoreset{equation}{section}
\makeatother

%%%%%%%%%%%%%%%%%%%%%%%%%%%%%%%%%%%%%%%%%%%%%%%%%%%%%%%%%%%%%%%%%
%                                                               %
%                NEW COMMANDS AND MACROS                        %
%                                                               %
%%%%%%%%%%%%%%%%%%%%%%%%%%%%%%%%%%%%%%%%%%%%%%%%%%%%%%%%%%%%%%%%%

%%%%% Simplify some frequently used LaTeX commands %%%%%

\nc{\be}{\begin{equation}}

\nc{\ee}{\end{equation}}

\nc{\bea}{\begin{eqnarray}}

\nc{\eea}{\end{eqnarray}}

\nc{\xx}{\nonumber\\}

\nc{\ct}{\cite}

\nc{\la}{\label}

\nc{\eq}[1]{(\ref{#1})}

\nc{\newcaption}[1]{\centerline{\parbox{6in}{\caption{#1}}}}

\nc{\fig}[3]{
\begin{figure}
\centerline{\epsfxsize=#1\epsfbox{#2.eps}}
\newcaption{#3. \label{#2}}
\end{figure}
}

%%% Caligraphic letters %%%%

\def\CA{{\cal A}}
\def\CC{{\cal C}}
\def\CD{{\cal D}}
\def\CE{{\cal E}}
\def\CF{{\cal F}}
\def\CG{{\cal G}}
\def\CH{{\cal H}}
\def\CK{{\cal K}}
\def\CL{{\cal L}}
\def\CM{{\cal M}}
\def\CN{{\cal N}}
\def\CO{{\cal O}}
\def\CP{{\cal P}}
\def\CR{{\cal R}}
\def\CS{{\cal S}}
\def\CU{{\cal U}}
\def\CV{{\cal V}}
\def\CW{{\cal W}}
\def\CY{{\cal Y}}
\def\CZ{{\cal Z}}

%%% Double line letters %%%

\def\IB{{\hbox{{\rm I}\kern-.2em\hbox{\rm B}}}}
\def\IC{\,\,{\hbox{{\rm I}\kern-.50em\hbox{\bf C}}}}
\def\ID{{\hbox{{\rm I}\kern-.2em\hbox{\rm D}}}}
\def\IF{{\hbox{{\rm I}\kern-.2em\hbox{\rm F}}}}
\def\IH{{\hbox{{\rm I}\kern-.2em\hbox{\rm H}}}}
\def\IN{{\hbox{{\rm I}\kern-.2em\hbox{\rm N}}}}
\def\IP{{\hbox{{\rm I}\kern-.2em\hbox{\rm P}}}}
\def\IR{{\hbox{{\rm I}\kern-.2em\hbox{\rm R}}}}
\def\IZ{{\hbox{{\rm Z}\kern-.4em\hbox{\rm Z}}}}

%%% Greek letters %%%

\def\a{\alpha}
\def\b{\beta}
\def\d{\delta}
\def\ep{\epsilon}
\def\ga{\gamma}
\def\k{\kappa}
\def\l{\lambda}
\def\s{\sigma}
\def\t{\theta}
\def\w{\omega}
\def\G{\Gamma}

%%%%% Mathematical Symbols

\def\half{\frac{1}{2}}
\def\dint#1#2{\int\limits_{#1}^{#2}}
\def\goto{\rightarrow}
\def\para{\parallel}
\def\brac#1{\langle #1 \rangle}
\def\curl{\nabla\times}
\def\div{\nabla\cdot}
\def\p{\partial}

%%%%% Roman pont in math

\def\Tr{{\rm Tr}\,}
\def\det{{\rm det}}

%%%%% Special Letters

\def\vare{\varepsilon}
\def\zbar{\bar{z}}
\def\wbar{\bar{w}}
%\def\what#1{\widehat{#1}}

%%%%% For this paper only

\def\ad{\dot{a}}
\def\bd{\dot{b}}
\def\cd{\dot{c}}
\def\dd{\dot{d}}
\def\so{SO(4)}
\def\bfr{{\bf R}}
\def\bfc{{\bf C}}
\def\bfz{{\bf Z}}

\begin{titlepage}

%---------------- preprint number ---------------

\hfill\parbox{3.7cm} {{\tt arXiv:2408.09653}}

\vspace{15mm}

\begin{center}
%------------------------ title ------------------------
{\Large \bf Explicit Construction of Hermitian Yang-Mills Instantons \\ on Coset Manifolds}

\vspace{10mm}
%---------------- authors and addresses ----------------

Jongmin Park \footnote{whdalsdl776@gist.ac.kr} and Hyun Seok Yang \footnote{hsyang@gist.ac.kr}
\\[10mm]

{\sl Department of Physics and Photon Science \\ Gwangju Institute of Science and Technology \\ Gwangju 61005, Korea}

\end{center}

\thispagestyle{empty}

\vskip1cm

%----------------------- abstract ----------------------

\centerline{\bf ABSTRACT}
\vskip 4mm
\noindent

In four dimensions, 't Hooft symbols offer a compact and powerful framework for describing
the self-dual structures fundamental to instanton physics.
Extending this to six dimensions, the six-dimensional ’t Hooft symbols can be constructed
using the isomorphism between the Lorentz group $Spin(6)$ and the unitary group $SU(4)$.
We demonstrate that the six-dimensional self-dual structures governed by the Hermitian Yang-Mills equations
can be elegantly organized using these generalized ’t Hooft symbols.
We also present a systematic method for constructing Hermitian Yang-Mills instantons
from spin connections on six-dimensional manifolds using the generalized 't Hooft symbols.
We provide a thorough analysis of the topological invariants such as instanton and Euler numbers.
\\

%PACS numbers: 11.10.Nx, 02.40.Gh, 11.25.Tq


\vspace{1cm}

\today

\end{titlepage}

\renewcommand{\thefootnote}{\arabic{footnote}}
\setcounter{footnote}{0}

\section{Introduction}

A six-dimensional gauge theory plays an important role in string theory.
Suppose that the ten-dimensional world is of the form $M^{3,1} \times M$, where $M^{3,1}$ is four-dimensional Minkowski spacetime and $M$ is a compact six-dimensional manifold. There is a restriction on possible compactifications of the string theory \cite{gsw-vol2}. An anomaly cancellation requires a subtle identity \cite{witten1984,classic1985}
\be \la{anomaly-string}
dH = \Tr R \wedge R - \Tr F \wedge F,
\ee
where $H$ is a globally defined three-form. Eq. \eq{anomaly-string} asserts that the cohomology classes represented by $\Tr R \wedge R$ and $\Tr F \wedge F$ must be equal. In general, the condition  \eq{anomaly-string} is very restrictive. One solution (or possibly only solution) is to set the gauge fields equal to the spin connections, suitably embedded in the gauge group of string theory, i.e. $SO(32)$ or $E_8 \times E_8$. On a manifold of $SU(3)$ holonomy such as a Calabi-Yau manifold  \cite{gsw-vol2},  the spin connection can be regarded as an $SU(3)$ gauge field. Therefore the anomaly constraint \eq{anomaly-string} may be solved by embedding the spin connection in the gauge group $SO(32)$ or $E_8 \times E_8$.

The six-dimensional Lorentz group $Spin(6)$ is isomorphic to the unitary group $SU(4)$, i.e.,
\be \la{spin6=su4}
Spin(6) \cong SU(4).
\ee
An eight-dimensional representation of $Spin(6)$, known as the spinor representation, is not irreducible. $Spin(6)$ has two irreducible spinor representations $S_\pm$, each of dimension $4$, namely the eigenspinors of the chiral operator $\Gamma_7 = i \Gamma^1 \cdots \Gamma^6$. The Lie algebra generators of the irreducible representation are given by $4 \times 4$ matrices
\be \la{spin6-chiral}
J^{AB}_\pm  \equiv \Gamma_\pm J^{AB}, \qquad (A, B = 1, \cdots, 6),
\ee
where $\Gamma_\pm = \frac{1}{2} (\mathbf{I}_8 \pm \Gamma_7)$ are projection operators defined by the chiral operator $\Gamma_7$ and $J^{AB} = \frac{1}{4}[\Gamma^A, \Gamma^B]$ are Lorentz generators of $Spin(6)$.  The group isomorphism \eq{spin6=su4} implies that the four-dimensional representation of $SU(4)$, known as the fundamental representation $\mathbf{4}$, has a similar split as the chiral spinor representation \cite{orai-book}. Indeed, reversing chirality corresponds to complex conjugation. Thus the complex conjugation exchanges the representation $\mathbf{4}$ with its complex conjugate $\overline{\mathbf{4}}$, called the anti-fundamental representation.
And the complex conjugate representation  $\overline{\mathbf{4}}$ is an inequivalent representation different from the original one $\mathbf{4}$.
The Lie algebra generators of the representations $\mathbf{4}$ and $\overline{\mathbf{4}}$ are $4 \times 4$ traceless (anti-)Hermitian matrices
\be \la{su4-matrix}
T^a_+ = T^a, \quad  T^a_- = (T^a)^*,  \qquad (a=1, \cdots, 15),
\ee
where $T^a \in su(4)$ are a four-dimensional generalization of Gell-Mann matrices.

Each basis of dimension 15 in \eq{spin6-chiral} and \eq{su4-matrix} forms a complete basis of $4 \times 4$ traceless (anti-)Hermitian matrices. Hence they are related by a linear transformation
\be \la{lin-tra}
J^{AB}_\pm = 2 \eta^{(\pm)a}_{AB} T^a_\pm   \quad \mathrm{or} \quad
T^a_\pm = \frac{1}{2} \eta^{(\pm)a}_{AB} J^{AB}_\pm.
\ee
The transformation coefficients in \eq{lin-tra} are determined by
\be \la{int-tsymbol}
\eta^{(\pm)a}_{AB} = - \Tr( T^a_\pm  J^{AB}_\pm ).
\ee
They correspond to a six-dimensional analogue of 't Hooft symbols in four dimensions, which provide a powerful tool in the context of non-Abelian gauge theories, in particular, instanton physics \cite{inst-book}.  The six-dimensional version of 't Hooft symbols was first introduced in \cite{yang-yun} (up to our best knowledge).  The 't Hooft symbols \eq{int-tsymbol} will also offer a significant advantage in tackling the group structure of gauge theories, especially, that arises in instanton calculations.

Mirror symmetry is a relationship between two topologically distinct Calabi-Yau manifolds that give rise to the exactly same physical theory in our world \cite{mirror-book}. It is a duality between two different Calabi-Yau manifolds $M$ and $\widetilde{M}$ such that the Hodge numbers of  $M$ and $\widetilde{M}$ satisfy the relations
\be \la{mirror-map}
h^{p,q} (M) = h^{3-p, q}( \widetilde{M}).
\ee
Since the Euler characteristic of a Calabi-Yau manifold $M$ is $\chi(M) = 2 \left( h^{1,1}(M) - h^{2,1}(M) \right)$, the mirror relation \eq{mirror-map} implies that
\be \la{mirror-euler}
\chi(M)  = - \chi(\widetilde{M}).
\ee
Therefore the most basic topology of two Calabi-Yau manifolds related by the mirror symmetry is different.
Nonetheless, the mirror symmetry implies that the type IIA string theory compactified on $M$
and the type IIB string theory compactified on $\widetilde{M}$ are dual descriptions of the same physics (see Chap. 19 in \cite{polchinski}).

We have the following motivations for this work (and its sequel):

1. In four dimensions, 't Hooft symbols provide a compact and efficient way to handle the self-dual structures
that arise in instanton physics \cite{inst-book}.
They also allow for a streamlined approach to identify self-dual gauge fields from spin connections on a four-manifold \cite{yang-op}.
An explicit construction was initiated by Charap and Duff \cite{char-duff1,char-duff2}
and further developed in \cite{oh-yang,lo-yang, psy-gist}.
We want to see how the six-dimensional 't Hooft symbols \eq{int-tsymbol} can be used to formalize the six-dimensional self-dual structures
and how to generalize the method in \cite{char-duff1,char-duff2}
to six-dimensional manifolds to find a solution of  Hermitian Yang-Mills (HYM) instantons.

2. In six dimensions, the second Chern number corresponding to the instanton number is always
positive (in the case with a trivial first Chern class) \cite{gang-tian,dryau}.
Thus, in six-dimensional space, there is no concept of an anti-instanton with a negative instanton number, unlike in four-dimensional space.
Instead, the third Chern number can take both signs.
It is known \cite{bott-tu} that the third Chern classes of a complex vector bundle $E$ and its conjugate bundle $\overline{E}$ have an opposite sign:
\be \la{3chern+-}
c_3 (E) = - c_3 ( \overline{E}).
\ee
The positive and negative chirality spinors of $Spin(6) \cong SU(4)$ transform as  $\mathbf{4}$ and $\overline{\mathbf{4}}$.
Then the Euler characteristic of $M$ is (see Eq. (4.2) in \cite{classic1985})
\bea \la{dirac-index}
\chi(M) &=& \mathrm{index} (\mathbf{4}) - \mathrm{index}(\overline{\mathbf{4}})  \xx
&=& \mathrm{index} (\mathbf{3}) - \mathrm{index}(\overline{\mathbf{3}}) = 2 \, \mathrm{index} (\mathbf{3}),
\eea
where $\mathrm{index} (R)$ is the Dirac index for chiral fermions in the $R$-representation.
When we combine the three statements, \eq{mirror-euler}, \eq{3chern+-} and \eq{dirac-index}, we get an idea \cite{yang-yun}
that the mirror structure of Calabi-Yau manifolds may be properly described by two kinds of 't Hooft symbols in \eq{int-tsymbol}
defined in the chiral and anti-chiral sectors.
We intend to establish a basic structure for this idea in this work and conduct a specific analysis in a follow-up study.

3. Recently, a model of cosmic inflation was constructed from the eight-dimensional Einstein-Yang-Mills theory
by putting Yang-Mills instantons on a four-dimensional internal space \cite{kky,kkt,hkky}.
What is interesting about this model is that the  cosmic inflation of the four-dimensional universe simultaneously
causes a dynamical compactification of the extra dimensions.
We want to extend this model to a ten-dimensional Einstein-Yang-Mills theory by constructing HYM instantons
on a six-dimensional internal space.
Then, we think that this inflationary model may be embedded in a ten-dimensional string theory.
A basic idea was speculated in Sect. 5 of Ref. \cite{kky}.

This paper is organized as follows.
In section 2, we recapitulate how HYM equations are derived from the Bogomolny inequality \cite{gang-tian,yang-yun}.
In section 3, we show that the six-dimensional self-dual structure described by the HYM equations
can be perfectly organized using the 't Hooft symbols.
In section 4, we explicitly construct HYM instantons
from the Fubini-Study metric of $\mathbb{C}\mathbb{P}^3$ using the 't Hooft symbols.
In section 5, we generalize the construction in section 4 to three coset manifolds
which are different realizations of $\mathbb{C}\mathbb{P}^3$.
These three coset spaces have been applied to the compactification of
ten-dimensional superstring theory \cite{dlust,mhs-coset,coset-comp1,coset-comp2} and $M$-theory \cite{atiyah-witten}.
First we show that the coset construction of $\mathbb{C}\mathbb{P}^3 = SU(4)/U(3)$ precisely reproduces
the coordinate representation in section 4.
Then we consider the other two cases.
The third example, a nonsymmetric $\mathbb{C}\mathbb{P}^3$ which is not K\"ahler but nearly K\"ahler \cite{jb-butr,toma2008},
shows very interesting properties after including a torsion \cite{dlust}.
In section 6, we compute the topological invariants of HYM instantons.
We suggest a new interpretation (or understanding) why the topological invariants of HYM instantons
derived from a gravitational solution have fractional numbers instead of integer numbers.
This symptom was attributed in \cite{gibb-pope} to a local nature of gauge connections.
We show that the contribution of the HYM instanton to the topological invariants is partial,
and that summing the contributions of other parts recovers integer topological numbers.
In the final section, we briefly summarize the results obtained and discuss some open issues.
In appendix A, we briefly review important properties of the 't Hooft symbols.
More details about the six-dimensional 't Hooft symbols \eq{int-tsymbol} can be found in Ref. \cite{yang-yun}.
Appendix B contains detailed results of the calculations that are necessary to obtain the main results
but are not presented in the main text.
In appendix C, we decompose the third Chern class into the part coming from HYM instantons
and the other parts to clarify why the topological invariant of the HYM instanton takes a fractional value.

\section{Gauge theory and Hermitian Yang-Mills instantons}

Suppose that the metric of a six-dimensional Riemannian manifold $M$ is given by
\begin{equation}\label{riem-metric}
  ds^2 = g_{MN}(x) dx^M dx^N = e^A \otimes e^A,
\end{equation}
where $e^A = e^A_M (x) dx^M, \; A =1, \cdots,6$, are an orthonormal basis
of one-forms on $M$. Let $\pi: E \to M$ be a vector bundle over $M$ with
a Lie group $G$ as its structure group. A connection $A$ on $E$ is defined by
specifying a covariant derivative
\begin{equation}\label{cov-deriv}
  D = d + A
\end{equation}
and the curvature of the connection $A$ is determined by $D^2$ as
\bea \label{curv-f}
F &=& dA + A \wedge A \nonumber \\
&=& \frac{1}{2} \left( \partial_M A_N - \partial_N A_M + [A_M, A_N] \right)
dx^M \wedge dx^N.
\eea
The Yang-Mills action on a six-dimensional Riemannian manifold $M$ is then given by
\begin{equation}\label{ym-action}
  S = - \frac{1}{2 g^2} \int_M d^6 x \sqrt{g} g^{MP} g^{NQ} \Tr F_{MN} F_{PQ}.
\end{equation}

One can rewrite the action \eq{ym-action} as \cite{yang-yun}
\bea \label{bogo-action}
  S &=& - \frac{1}{4 g^2} \int_M d^6 x \sqrt{g}  \Tr \left[
  \left( F_{AB} \pm \frac{1}{4} \varepsilon^{ABCDEF}
  F_{CD} I_{EF} \right)^2 - \frac{1}{2} \left(I_{AB} F^{AB} \right)^2 \right. \xx
  && \left. \hspace{4cm} \mp \frac{1}{2} \varepsilon^{ABCDEF} F_{AB} F_{CD} I_{EF} \right],
\eea
where $I = \textbf{I}_3 \otimes i \sigma^2$ is a $6 \times 6$ antisymmetric matrix
of rank 6. The action \eq{bogo-action} may be written in a more succinct form
\bea \label{bogomolny}
  S &=& - \frac{1}{4 g^2} \int_M d^6 x \sqrt{g}  \Tr \left[
  \left( F_{AB} \pm *(F \wedge \Omega)_{AB} \right)^2 - \frac{1}{2} \left(I_{AB} F^{AB} \right)^2 \right] \xx
  && \pm \frac{1}{g^2} \int_M \Tr F \wedge F \wedge \Omega,
\eea
where $\Omega = \frac{1}{2} I_{AB} e^A \wedge e^B$ is the two-form of maximal rank and
\begin{equation}\label{proj-1}
  *(F \wedge \Omega)_{AB} \equiv \frac{1}{4} \varepsilon^{ABCDEF} F_{CD} I_{EF}.
\end{equation}
Using the fact
$$ \Tr F \wedge F = d \Tr \left(A \wedge F - \frac{1}{3} A \wedge A \wedge A \right) \equiv dK, $$
one can see that the last term in Eq. \eq{bogomolny} is a topological term, i.e.,
\begin{equation}\label{top-term}
\Tr F \wedge F \wedge \Omega =  d (K \wedge \Omega)
\end{equation}
if the two-form $\Omega$ is closed, i.e., $d\Omega = 0$.

If the condition \eq{top-term} is obeyed, the minimum of the action \eq{bogomolny} is achieved
in a configuration satisfying the equations \cite{cdfn,gang-tian}
\begin{equation}\label{hym-eq}
F_{AB} = \mp  *(F \wedge \Omega)_{AB}.
\end{equation}
In terms of local coordinates, Eq. \eq{hym-eq} can be written as
\begin{equation}\label{local-hym}
F_{MN} = \mp \frac{1}{4} \frac{\varepsilon^{KLPQRS}}{\sqrt{g}}
g_{MK} g_{NL} F_{PQ} \Omega_{RS}.
\end{equation}
It is important to note that the Bogomolny equations \eq{hym-eq} directly lead to the following equation
\begin{equation}\label{stable-eq}
I_{AB} F^{AB} = 0 \quad \cong \quad F \wedge \Omega^2 = 0,
\end{equation}
which can be checked using the identity $\frac{1}{8} \varepsilon^{ABCDEF} I_{CD} I_{EF}
= I_{AB}$.
It is straightforward to show that a solution of the Bogomolny equations \eq{hym-eq}
automatically satisfies the Yang-Mills equations of motion
\begin{equation}\label{ym-eom}
g^{MN} D_M F_{NP} = 0.
\end{equation}

The Hermitian Yang-Mills (HYM) equation \eq{hym-eq} is a generalization of the self-duality equation in four dimensions
to six-dimensional manifolds. Such equations in higher dimensions were first introduced in \cite{cdfn}.
See also \cite{bakasi} for a discussion from topological quantum field theories in various dimensions and
\cite{blp-hbps} for higher-dimensional BPS equations and their generalization to noncommutative spaces.
There also has been an extensive study finding solutions ($G$- or $H$-connections) on group manifolds
and coset manifolds $G/H$ \cite{lech-popo1,lech-popo2,lech-popo3,lech-popo4,lech-popo5}.
We will consider a different method to find different solutions of Eq. \eq{hym-eq}.
We will directly derive $SU(3)$ connections satisfying Eq. \eq{hym-eq} from the spin connections on a six-dimensional manifold
using the method in \cite{char-duff1,char-duff2},
where the 't Hooft symbols \eq{int-tsymbol} will be of great help.

We showed in appendix B that the space of two-forms $\Omega^2_\pm(M)$
on any six-dimensional manifold $M$ can be decomposed into three subspaces
\be \la{2form-decomp}
\Omega^2_\pm(M) = \Omega^2_{\pm 1} \oplus \Omega^2_{\pm 6} \oplus \Omega^2_{\pm 8}.
\ee
This decomposition can also be understood by defining a $\Omega$-Hodge-dual operator acting
on $\Omega^2_\pm (M)$ \cite{gang-tian,yang-yun}
\begin{equation}\label{w-hodge}
    *_{\Omega}: \Omega^2_\pm (M) \to \Omega^2_\pm (M)
\end{equation}
by $*_{\Omega} (\alpha) = *(\alpha \wedge \Omega)$ for $\alpha \in \Omega^2_\pm (M)$.
One can show \cite{cdfn,yang-yun} that the three subspaces in \eq{2form-decomp} are the eigenspaces of
the operator $*_{\Omega}$ of dimension $1, 6$ and $8$ with eigenvalues $2, 1$ and $-1$, respectively.
Thus the solution satisfying  the $\Omega$-anti-self-dual equations
($-$-sign in \eq{hym-eq}) belongs to the space $\Omega^2_{\pm 8}$ whereas the solution satisfying  the $\Omega$-self-dual equations ($+$-sign in \eq{hym-eq}) belongs
to the space $\Omega^2_{\pm 6}$. In particular, the solution satisfying
the $\Omega$-anti-self-dual equations in \eq{hym-eq} is called HYM instantons.
The result \eq{iu3} implies that the HYM instantons take values
in $su(3)$ Lie algebra.

From the derivation of the HYM equations \eq{hym-eq},
one can see that, in order to define a HYM instanton on a six-dimensional manifold $M$,
it is necessary for the manifold $M$ to have a closed two-form $\Omega$.
A K\"ahler manifold is exactly such a case where $\Omega$ is the K\"ahler form of
the manifold. In this case, the curvature two-form $F$ can be decomposed as
\begin{equation} \label{holo-decom}
F =  F^{(2,0)} + F^{(1,1)} + F^{(0,2)}
\end{equation}
according to the complex structure of the K\"ahler manifold $M$.
Then the HYM equations \eq{hym-eq} can be stated as the form
\be \label{duy}
F^{(2,0)} = F^{(0,2)} = 0, \qquad F^{(1,1)} \cdot \Omega = 0,
\ee
where $F^{(1,1)} \cdot \Omega$ denotes the orthogonal projection of $F^{(1,1)}$
in the $\Omega$-direction. The first equation in \eq{duy} states that its solution
is a holomorphic vector bundle and the second equation requires
that the holomorphic vector bundle must be a `stable' bundle (see Sect. 15.6 in \cite{gsw-vol2}).
Therefore, the HYM instantons are in one-to-one correspondence
with stable holomorphic bundles over a K\"ahler manifold.

The equations \eq{duy} are also known as the Donaldson-Uhlenbeck-Yau (DUY) equation for the moduli space
of stable holomorphic vector bundles on a K\"ahler manifold \cite{duy1,duy2}.
They also appear in the Calabi-Yau compactification of the heterotic strings where
Eq. \eq{duy} is the condition for the existence of an unbroken supersymmetry \cite{gsw-vol2}.

However it appears that HYM instantons can be defined on a six-dimensional
manifold $M$ as long as $M$ admits a non-degenerate closed two-form $\Omega$
though $M$ is not necessarily K\"ahler.
We will study to what extent our method can be extended.

\section{Hermitian Yang-Mills instantons from K\"ahler manifolds}

The HYM equations \eq{hym-eq} are first-order nonlinear differential equations
of a $G$-connection $A$ on a K\"ahler manifold.
Hence it is generally difficult to find a solution of Eq. \eq{hym-eq}
on a general K\"ahler manifold. Now we will introduce a general method
for finding a solution of the HYM equations \eq{hym-eq}, given a six-dimensional manifold $M$.
A basic idea is to identify a vector bundle $E$ over $M$ with the tangent bundle $TM$ of $M$.
For this purpose, we take the gauge group $G$ as $SU(4) \cong Spin(6)$.
An important step is to separate the components of $\Omega^2_{\pm 8}$
from the space $\Omega^2_\pm(M)$ in \eq{2form-decomp}.
The 't Hooft symbols constructed in appendix A will play an important role.

Using the metric \eq{riem-metric}, one can determine the spin connection ${\omega^A}_B ={{\omega_M}^A}_B dx^M$
and curvature tensors ${R^A}_B = \frac{1}{2}{{R_{MN}}^A}_B dx^M \wedge dx^N$
by solving the structure equations \cite{egh-report,nakahara}
\bea \label{steq-torsion}
&& T^A = de^A + {\omega^A}_B \wedge e^B = 0, \\
\label{steq-curvature}
&& {R^A}_B = d{\omega^A}_B + {\omega^A}_C \wedge {\omega^C}_B.
\eea
Define $su(4)$ Lie algebra generators
\be \la{6dim-su4}
\mathcal{T}^{(\pm)a}_{AB} \equiv [\eta^{(\pm)a}]_{AB}.
\ee
They provide an antisymmetric 6-dimensinal representation of $SU(4)$ and satisfy the commutation relation
\begin{equation} \label{6dim-comm}
  [\mathcal{T}^{(\pm)a}, \mathcal{T}^{(\pm)b}] = - f^{abc} \mathcal{T}^{(\pm)c},
\end{equation}
which can be derived by using the relations Eqs. (B.11) and (B.12) in \cite{yang-yun}.

We identify $SU(4)$ gauge fields as
\be  \label{su4-gauge-id}
A^{(\pm) a} = \omega_{AB} \eta^{(\pm)a}_{AB}
\quad \mathrm{or} \quad \omega_{AB} = A^{(\pm) a} \eta^{(\pm)a}_{AB}.
\ee
Note that the vector space of the Lorentz generators in \eq{chiral-gen} is isomorphic
to the vector space of two-forms $\Omega_\pm^2 (M)$.
Therefore the decomposition \eq{2form-decomp} can also be applied
to the decomposition of $SU(4)$ gauge fields \cite{yang-yun}:
\be  \label{su4-gauge-decom}
A^{(\pm) a} = A^{(\pm) 0} + A^{(\pm) \dot{a}} + A^{(\pm) \hat{a}},
\ee
where
\be \la{a0}
A^{(\pm) 0} \equiv \pm A^{(\pm)3} + \frac{1}{\sqrt{3}} A^{(\pm)8}
+ \frac{1}{\sqrt{6}} A^{(\pm)15}
\ee
is a $U(1)$ gauge field and $A^{(\pm) \dot{a}}$ and $A^{(\pm) \hat{a}}$ are gauge fields
along the direction of Lie algebra generators obeying \eq{icoset} and \eq{iu3}, respectively.
In other words, $A^{(\pm) \dot{a}}$ are gauge fields in the coset $\mathbb{C}\mathbb{P}^3 = SU(4)/U(3)$
and $A^{(\pm) \hat{a}}$ are $SU(3)$ gauge fields.\footnote{In particular, we will call
the coset space $\mathbb{C}\mathbb{P}^3 = SU(4)/U(3)$ or the subspace $\Omega^2_{\pm 6}$
as the dotted space (or part) according to this notation.}
The entries in $A^{(\pm) \dot{a}}$ and $A^{(\pm) \hat{a}}$ coincide with those in $\Omega^2_{\pm 6}$
and $\Omega^2_{\pm 8}$, respectively,
in Eq. \eq{168-decom}. As a result, $(A^{(\pm) 0}, A^{(\pm) \hat{a}})$ form $U(3)$ gauge fields.

Riemann curvature tensors in \eq{steq-curvature} are similarly mapped
to $SU(4)$ gauge fields:
\bea  \label{su4-field-id}
F^{(\pm) a} &=& \eta^{(\pm)a}_{AB} R_{AB} \xx
&=& dA^{(\pm) a} - \frac{1}{2} f^{abc} A^{(\pm) b} \wedge A^{(\pm) c}.
\eea
Since the field strengths in \eq{su4-field-id} are $su(4)$-valued two-forms,
they can be expanded in the two-form basis \eq{2-forms} as
\be \label{exp-f}
F^{(\pm) a} = f^{ab}_{(\pm\pm)} J^b_\pm.
\ee
The coefficients $f^{ab}_{(\pm\pm)}$ satisfy the following properties \cite{yang-yun}
\be \label{coeff-symm}
 f^{ab}_{(\pm\pm)} = f^{ba}_{(\pm\pm)}, \\\qquad
 d^{abc}f^{bc}_{(\pm\pm)} = 0,
\ee
where $d^{abc}$ are symmetric structure constants of $SU(4)$.
They are derived from the Bianchi identity of curvature tensors:
$R_{ABCD} + R_{ACDB} + R_{ADBC} = 0$.

Now we impose the K\"ahler condition, $d \Omega = 0$.
It can be recast as the form
\bea \la{kahler-cond}
0 = d \Omega &=& - \left (I_{AB} {\omega^B}_C \right) \wedge e^A \wedge e^C \xx
&=& - \frac{1}{2} \left( I_{AB} \eta^{(\pm)a}_{BC} - \eta^{(\pm)a}_{AB} I_{BC} \right) A^{(\pm)a} \wedge e^A \wedge e^C \\
&=& - \frac{1}{2} \left(  [I, \mathcal{T}^{(\pm)a}] \right)_{AC}
A^{(\pm)a} \wedge e^A \wedge e^C, \nonumber
\eea
where we used the torsion-free condition \eq{steq-torsion} and
the identifications in \eq{6dim-su4} and \eq{su4-gauge-id}.
Therefore, the K\"ahler condition is equivalent to the condition \eq{iu3},
i.e. $A^{(\pm) \dot{a}} = 0$ in the decomposition \eq{su4-gauge-decom}.
In the end, after imposing the K\"ahler condition, the gauge fields in \eq{su4-gauge-decom} must be $u(3)$-valued, i.e.,
\be  \label{kahler-gauge}
A^{(\pm) a} = A^{(\pm) 0} + A^{(\pm) \hat{a}}.
\ee
Since $A^{(\pm) 0}$ are $U(1)$ gauge fields, they are not mixed with
$SU(3)$ gauge fields $A^{(\pm) \hat{a}}$ so that
\be  \label{kahler-field}
F^{(\pm) a} = F^{(\pm) 0} + F^{(\pm) \hat{a}},
\ee
where $F^{(\pm) 0} = d A^{(\pm) 0}$ and $F^{(\pm) \hat{a}}
= dA^{(\pm) \hat{a}} - \frac{1}{2} f^{\hat{a}\hat{b}\hat{c}} A^{(\pm) \hat{b}} \wedge A^{(\pm) \hat{c}}$.

The $U(1)$ field strength $F^{(\pm) 0} = d A^{(\pm) 0}$ is related to
the Ricci curvature $R = i R_{\alpha \overline{\beta}} dz^\alpha \wedge d \overline{z}^{\bar{\beta}}
= i \partial \overline{\partial} \ln \det g_{\alpha \overline{\beta}}$ and it identically vanishes for a Ricci-flat manifold.
Then the expansion \eq{exp-f} for the $SU(3)$ field strengths $F^{(\pm) \hat{a}}$ reduces to
\be \label{exp-su3f}
F^{(\pm) \hat{a}} = f^{\hat{a}\hat{b}}_{(\pm\pm)} J^{\hat{b}}_\pm,
\ee
where $J^{\hat{b}}_\pm \in \Omega^2_{\pm 8}$.
Therefore they automatically satisfy the HYM equations
\be \la{hym-su3}
F^{(\pm) \hat{a}} = - * \left( F^{(\pm) \hat{a}} \wedge \Omega \right),
\qquad \hat{a} \in su(3).
\ee
So we have succeeded in constructing an instanton solution satisfying
the HYM equations defined on a general K\"ahler manifold.
We will apply this method to specific K\"ahler manifolds in the next section.

\section{Hermitian Yang-Mills instantons on $\mathbb{C}\mathbb{P}^3$}

In this section, we will illustrate how to apply the method in the previous section
to a well-known K\"ahler manifold, $\mathbb{C}\mathbb{P}^3$.
We take the Fubini-Study metric on $\mathbb{C}\mathbb{P}^3$ considered
in Ref. \cite{aru-fro}:
\be \la{cp3-metric}
ds^2 = d\rho^2 + \frac{1}{4} \sin^2 2 \rho \left(d\phi
+ \frac{1}{2} \sin^2 \theta (d \alpha_3 + \cos \alpha_1 d\alpha_2) \right)^2
+ \sin^2 \rho \, ds^2_{\mathbb{C}\mathbb{P}^2},
\ee
where
\be \la{cp2-metric}
ds^2_{\mathbb{C}\mathbb{P}^2} = d \theta^2 + \frac{1}{4} \sin^2 \theta \left(d\alpha^2_1
+ \sin^2 \alpha_1 d \alpha_2^2 + \cos^2 \theta ( d\alpha_3 + \cos \alpha_1 d\alpha_2)^2 \right)
\ee
is the Fubini-Study metric on $\mathbb{C}\mathbb{P}^2$.
We choose an orthonormal basis of one-forms $e^A, \; A=1, \cdots, 6$,
\bea \la{vielbein}
&& e^1 = d\rho, \quad e^2 = \frac{1}{2} \sin 2 \rho
\left( d \phi + \frac{1}{2} \sin^2 \theta
(d \alpha_3 + \cos \alpha_1 d\alpha_2 ) \right),
\quad e^3 =  \sin \rho d \theta, \\
&& e^4 = \frac{1}{2} \sin \rho \sin \theta \cos \theta
( d\alpha_3 + \cos \alpha_1 d\alpha_2),
\quad e^5 = \frac{1}{2} \sin \rho \sin \theta \sin \alpha_1 d \alpha_2,
\quad e^6 = \frac{1}{2} \sin \rho \sin \theta d \alpha_1. \nonumber
\eea
It is straightforward to determine spin connections and curvature tensors
for the metric \eq{cp3-metric} by solving the structure equations
\eq{steq-torsion} and \eq{steq-curvature}. Nonvanishing components are listed
in appendix B.

Using the explicit results in appendix B.1, it is straightforward to identify
the $su(3) \subset su(4)$ gauge fields and their field strengths
(we follow Ref. \cite{yang-yun} for the ordering of $su(3)$ gauge fields,
$\hat{a} = 1, \cdots, 8$)
\bea \la{su3-gauge}
&& A^{(+) \hat{1}} = A^{(+)13} = - 2 \cot \rho e^4,
\quad A^{(+) \hat{2}} = A^{(+)14} = - 2 \cot \rho e^3, \xx
&& A^{(+) \hat{3}} = \frac{1}{\sqrt{3}} ( A^{(+)8} - \sqrt{2} A^{(+)15} )
= - \tan \rho e^2 - 2 \csc \rho \cot 2 \theta e^4, \xx
&& A^{(+) \hat{4}} = A^{(+)6} = - 2 \cot \rho e^6,
\quad A^{(+) \hat{5}} = A^{(+)7} = 2 \cot \rho e^5, \xx
&& A^{(+) \hat{6}} = A^{(+)11} = - 2 \csc \rho \cot \theta e^6,
\quad A^{(+) \hat{7}} = A^{(+)12} = 2 \csc \rho \cot \theta e^5, \xx
&& A^{(+) \hat{8}} = \frac{2}{\sqrt{3}} \left( - \frac{1}{2} A^{(+)3}
+ \frac{1}{\sqrt{3}} A^{(+)8} + \frac{1}{\sqrt{6}} A^{(+)15} \right) \xx
&& \hspace{1cm} = \frac{1}{\sqrt{3}} ( - \tan \rho e^2
- 2 \csc \rho \csc 2 \theta e^4 + 4 \csc \rho \csc \theta \cot \alpha_1 e^5 ),
\eea
and
\bea \la{su3-field}
&& F^{(+) \hat{1}} = F^{(+)13}
= 2 e^1 \wedge e^4 - 2 e^2 \wedge e^3, \quad
F^{(+) \hat{2}} = F^{(+) 14} = 2 e^1 \wedge e^3 + 2 e^2 \wedge e^4,  \xx
&& F^{(+) \hat{3}} = \frac{1}{\sqrt{3}} ( F^{(+)8} - \sqrt{2} F^{(+)15} )
= - 2 e^1 \wedge e^2 + 2 e^3 \wedge e^4, \xx
&& F^{(+) \hat{4}} = F^{(+) 6} = 2 e^1 \wedge e^6 - 2 e^2 \wedge e^5,
\quad F^{(+) \hat{5}} = F^{(+) 7} = - 2 e^1 \wedge e^5 - 2 e^2 \wedge e^6, \xx
&& F^{(+) \hat{6}} = F^{(+) 11} = 2 e^3 \wedge e^6 - 2 e^4 \wedge e^5, \quad
F^{(+) \hat{7}} = F^{(+) 12} = - 2 e^3 \wedge e^5 - 2 e^4 \wedge e^6, \xx
&& F^{(+) \hat{8}} = \frac{2}{\sqrt{3}} \left( - \frac{1}{2} F^{(+)3}
+ \frac{1}{\sqrt{3}} F^{(+)8} + \frac{1}{\sqrt{6}} F^{(+)15} \right) \xx
&& \hspace{1cm} = \frac{2}{\sqrt{3}} ( - e^1 \wedge e^2 - e^3 \wedge e^4
+ 2 e^5 \wedge e^6).
\eea
The curvature two-forms in \eq{su3-field} precisely take the form \eq{exp-su3f},
i.e. $F^{(+) \hat{a}} = 2  J^{\hat{a}}_+ \in \Omega^2_{+8}$.
We also get exactly the same result for the anti-chiral sector
using Eqs. \eq{-su4-gauge} and \eq{-su4-field}, i.e.,
$F^{(+) \hat{a}} = F^{(-) \hat{a}}$, although their $SU(4)$ gauge fields
are slightly different from the chiral sector \eq{su4-gauge}.

Since $\mathbb{C}\mathbb{P}^3$ is a K\"ahler manifold, gauge fields
$A^{(\pm) \dot{a}} \; (\dot{a} = 1, \dots, 6)$
in the dotted space must be trivial as was proved in \eq{kahler-cond}.
It was shown in appendix B that it is the case.
Since $\mathbb{C}\mathbb{P}^3$ is an Einstein-K\"ahler manifold
with the Ricci-scalar $R = 48$ (where the radius is normalized to be $1$),
it must have a nontrivial $U(1)$ gauge field.
We found that
\bea \la{u1-field}
A^{(\pm) 0} &=& \pm A^{(\pm)3} + \frac{1}{\sqrt{3}} A^{(\pm)8}
+ \frac{1}{\sqrt{6}} A^{(\pm)15} \xx
&=& (\tan \rho e^2 - 3 \cot \rho ) e^2 + \csc \rho ( \tan \theta
- 2 \cot \theta) e^4 + 2 \csc \rho \csc \theta \cot \alpha_1 e^5, \xx
F^{(\pm) 0} &=& d A^{(\pm) 0} = 8 \Omega.
\eea
It is straightforward to check that the field strengths $F^{(\pm) \hat{a}}$
in \eq{su3-field} obey the HYM equations \eq{hym-su3}.
Therefore, $SU(3)$ gauge fields in \eq{su3-gauge} (locally) describe a
HYM instanton on $\mathbb{C}\mathbb{P}^3$.

\section{Coset construction}

$\mathbb{C}\mathbb{P}^3$ admits two different homogeneous Einstein metrics.
The usual Fubini-Study metric has $SU(4)$ symmetry which can be viewed as the homogeneous space $SU(4)/U(3)$,
while the second one is invariant under the subgroup
of $SU(4)$ that can be usefully viewed as the homogeneous space $Sp(4)/SU(2) \times U(1)$.
The latter may be regarded as an $\mathbb{S}^2$-bunble over $\mathbb{S}^4$ (so the twistor space of $\mathbb{S}^4$
\cite{atiyah-1979})
and it also allows a non-symmetric coset space \cite{dlust,mhs-coset}.
The coset space $SU(4)/U(3)$ emphasizes the complex structure of $\mathbb{C}\mathbb{P}^3$,
so it may be a natural choice for problems involving complex geometry.
But the coset space $Sp(4)/SU(2) \times U(1)$ highlights the symplectic structure of $\mathbb{C}\mathbb{P}^3$
which may be useful in the context where symplectic geometry is relevant.
We will compare HYM instantons on two different coset spaces;
$SU(4)/U(3)$ vs. $Sp(4)/SU(2) \times U(1)$.

Consider a compact semi-simple Lie group $G$ and a closed subgroup $H$ of $G$ such that
the coset space $G/H$ is a reductive homogeneous space.
See \cite{salam-coset,kazo-report} for a review of coset spaces.
Split the Lie algebra of $G$ as $\mathfrak{g} = \mathfrak{h} \oplus \mathfrak{k}$, where
$\mathfrak{h} = \{ \mathcal{H}_{\bar{a}} | \bar{a} = 1, \cdots, \mathrm{dim} (H) \}$ denotes
the Lie algebra of $H$ and $\mathfrak{k} = \{ \mathcal{K}_i | i = 1, \cdots, \mathrm{dim} (G) - \mathrm{dim} (H) \}$
contains the coset generators.
The structure constants of $G$ are defined by the commutation relations
\bea \la{coset-comm}
&& [\mathcal{H}_{\bar{a}}, \mathcal{H}_{\bar{b}} ] = {C_{\bar{a} \bar{b}}}^{\bar{c}} \mathcal{H}_{\bar{c}}, \xx
&& [\mathcal{H}_{\bar{a}}, \mathcal{K}_i ] = {C_{\bar{a} i}}^{\bar{b}} \mathcal{H}_{\bar{b}} + {C_{\bar{a} i}}^j \mathcal{K}_j, \\
&& [\mathcal{K}_i, \mathcal{K}_j ] = {C_{ij}}^{\bar{a}} \mathcal{H}_{\bar{a}}
+ {C_{ij}}^k \mathcal{K}_k. \nonumber
\eea
Since we have assumed that $G/H$ is reductive, ${C_{\bar{a} i}}^{\bar{b}} = 0$.
For a symmetric coset space, ${C_{ij}}^k = 0$.

Let $x$ be local coordinates on $G/H$ and choose a coset representative $L_x$ from
each coset. The decomposition $\mathfrak{g} = \mathfrak{h} \oplus \mathfrak{k}$
of the Lie algebra of $G$ leads to a corresponding decomposition of the one-form
\be \la{mc-form}
\Theta = L_x^{-1} d L_x = \omega^{\bar{a}} \mathcal{H}_{\bar{a}} + \theta^i \mathcal{K}_i.
\ee
The one-form $\Theta$ satisfies the Cartan-Maurer equation
\be \la{cm-eq}
d\Theta + \Theta \wedge \Theta = 0,
\ee
which, using \eq{coset-comm}, yields \cite{salam-coset,kazo-report}
\bea \la{comp-cm}
&& d \theta^i = - \frac{1}{2} {C_{jk}}^i \theta^j \wedge \theta^k
- {C_{\bar{a}j}}^i \omega^{\bar{a}} \wedge \theta^j, \xx
&& d \omega^{\bar{a}} = - \frac{1}{2} {C_{\bar{b} \bar{c}}}^{\bar{a}}
\omega^{\bar{b}} \wedge \omega^{\bar{c}} - \frac{1}{2} {C_{ij}}^{\bar{a}} \theta^i \wedge \theta^j.
\eea
Any metric on $G/H$ can be written in terms of the coset coframe $\theta^i$ as
\be \la{coset-metric}
ds^2 = g_{ij} \theta^i \otimes \theta^j,
\ee
where $g_{ij}$ is a $G$-invariant metric and constant by homogeneity.
The Levi-Civita connection one-form ${\omega^i}_j$ of a metric $g$ should satisfy
$\omega_{ij} \equiv g_{ik} {\omega^k}_j = - \omega_{ji}$ and they are uniquely determined
by the torsion-free condition \cite{mhs-coset}
\be  \la{coset-tor}
d \theta^i + {\omega^i}_j \wedge \theta^j = 0.
\ee
A general solution of the torsion-free condition \eq{coset-tor} is found to be \cite{mhs-coset}
\be \la{levi-conn}
{\omega^i}_j = {C_{\bar{a}j}}^i \omega^{\bar{a}} + {D_{kj}}^i \theta^k
\ee
with
\begin{eqnarray} \la{coeff-dk}
&& {D_{kj}}^i = \frac{1}{2} g^{im} \left( {C_{mj}}^l g_{lk}
+ {C_{kj}}^l g_{ml} - {C_{km}}^l g_{jl} \right)
= \frac{1}{2} {C_{kj}}^i + {K_{kj}}^i, \xx
&&  {K_{kj}}^i = \frac{1}{2} g^{im} \left( {C_{mk}}^l g_{jl}
+ {C_{mj}}^l g_{kl} \right).
\end{eqnarray}
Then the curvature two-form ${R^i}_j = d{\omega^i}_j + {\omega^i}_k \wedge {\omega^k}_j$ is determined as follows:
\be \la{coset-curv}
{R^i}_j = \frac{1}{2} \left( {D_{lj}}^m {D_{km}}^i - {D_{kj}}^m {D_{lm}}^i
- {C_{kl}}^{\bar{a}} {C_{\bar{a}j}}^i - {C_{kl}}^m {D_{mj}}^i \right)
\theta^k \wedge \theta^l.
\ee

Now we will apply the coset construction to find HYM instantons
on the homogeneous spaces $SU(4)/U(3)$ and $Sp(4)/SU(2) \times U(1)$.
These coset spaces describe $\mathbb{C}\mathbb{P}^3$ but admit different homogeneous Einstein metrics.

\subsection{$\mathbb{C}\mathbb{P}^3 = SU(4)/SU(3) \times U(1)$}

We choose the following basis of the Lie algebra $\mathfrak{g}$ of $G=SU(4)$:
\bea \la{su4-cobasis}
 && \mathcal{H}_{\bar{a}} = 2 T_{\bar{a}}, \qquad  {\bar{a}} = 1, \cdots, 8, 15, \xx
 && \mathcal{K}_i = 2 T_{8 + i}, \qquad  i = 1, \cdots, 6,
\eea
where $T_a = (T_{\bar{a}}, \; T_{8 + i})$ are Lie algebra generators used
in \eq{iden-liealg}. The subgroup $H = SU(3) \times U(1)$ is generated by
$2 T_{\bar{a}} \; ({\bar{a}} = 1, \cdots, 8, 15)$.
Thus the structure constants in \eq{coset-comm} are given by
\be \la{su4-cos-con}
{C_{\bar{a} \bar{b}}}^{\bar{c}} = - 2 {f_{\bar{a} \bar{b}}}^{\bar{c}}, \quad
{C_{\bar{a} i}}^{j} = - 2 {f_{\bar{a} (8+i)}}^{(8+j)}, \quad
{C_{ij}}^{\bar{a}} = - 2 {f_{(8+i)(8+j)}}^{\bar{a}}.
\ee

The coset space $G/H$ is symmetric, so ${C_{ij}}^{k} = 0$.
In this case, the connection one-form takes a simple form
\be \la{co-cps-conn}
{\omega^i}_j  = - 2 {f_{\bar{a} (8+j)}}^{(8+i)} \omega^{\bar{a}}.
\ee
Their entries are listed in appendix B.2.
Using the results in appendix B.2, one can easily find the $su(3) \subset su(4)$ gauge fields
obeying the HYM equations (3.13):
\bea \la{cos-su3-conn}
&& A^{(+) \hat{1}} = A^{(+)13} = 2 \omega^{1},
\quad A^{(+) \hat{2}} = A^{(+)14} = 2  \omega^{2}, \quad
A^{(+) \hat{3}} = \frac{1}{\sqrt{3}} ( A^{(+)8} - \sqrt{2} A^{(+)15} )
= - 2  \omega^{3},  \xx
&& A^{(+) \hat{4}} = A^{(+)6} =  2 \omega^{4},
\quad A^{(+) \hat{5}} = A^{(+)7} =  - 2 \omega^{5},
\quad A^{(+) \hat{6}} = A^{(+)11} = 2  \omega^{6}, \\
&& A^{(+) \hat{7}} = A^{(+)12} = - 2  \omega^{7},
\quad A^{(+) \hat{8}} = \frac{2}{\sqrt{3}} \left( - \frac{1}{2} A^{(+)3}
+ \frac{1}{\sqrt{3}} A^{(+)8} + \frac{1}{\sqrt{6}} A^{(+)15} \right) =
- 2 \omega^{8}, \nonumber
\eea
and
\be \la{cos-a0}
A^{(+)0} = A^{(+)3} + \frac{1}{\sqrt{3}} A^{(+)8}
+ \frac{1}{\sqrt{6}} A^{(+)15} = 6 \sqrt{\frac{2}{3}} \omega^{15}.
\ee
Their curvature two-forms take exactly the same expression as \eq{su3-field}
after the identification $\theta^i \cong e^A$.
Therefore, $su(3)$-connections in \eq{cos-su3-conn} describe a HYM
instanton on $\mathbb{C}\mathbb{P}^3$ obeying Eq. (3.13).
These instanton connections can equally be embedded into the anti-chiral sector
with the same structure.

\subsection{Symmetric $Sp(4)/[SU(2) \times U(1)]_{max}$}

The homogeneous space $Sp(4)/[SU(2) \times U(1)]_{max} \cong \mathbb{C}\mathbb{P}^3$ can be viewed
as a twistor space of 4-dimensional symmetric space $\mathbb{S}^4$ \cite{atiyah-1979} and the properties
of this coset space are very different from the previous example.
The twistor space $Z$ of a four-dimensional Riemannian manifold is equipped with
two natural almost complex structures. See sect. 3 in \cite{jb-butr}.
The first one $(Z, g_1, J_+)$ is K\"ahler but the second one $(Z, g_2, J_-)$
is nearly K\"ahler. The homogeneous space $M = Sp(4)/[SU(2) \times U(1)]_{max}$ corresponds
to the second case. (See \textbf{Proposition 3.2} in \cite{jb-butr} and Eq. (3.6) in \cite{toma2008}.)
Thus we cannot use $J_-$ to define a closed two-form.
But there exists an $Sp(4)$-invariant closed two-form on $M$ \cite{mhs-coset}.
Then we can define the HYM equations \eq{hym-su3}
using the $Sp(4)$-invariant closed two-form although $M$ is not a K\"ahler manifold.
We will show this claim with the explicit example.

We choose the following basis of the Lie algebra in the fundamental representation of $Sp(4)$:
\bea \la{lie-sp4}
T_{2{\bar{i}}-1} = \frac{i}{2} (\sigma^3 \otimes \sigma^{\bar{i}}) = \frac{i}{2} \left( \begin{array}{cc}
            \sigma^{\bar{i}}  & 0 \\
            0 & - \sigma^{\bar{i}} \\
          \end{array} \right), \quad
T_{2{\bar{i}}} = \frac{i}{2} (\sigma^1 \otimes \sigma^{\bar{i}}) = \frac{i}{2} \left( \begin{array}{cc}
            0 & \sigma^{\bar{i}}  \\
            \sigma^{\bar{i}}  & 0 \\
          \end{array} \right), \xx
T_{6 + \bar{i}} = \frac{i}{2} (\mathbf{I}_2 \otimes \sigma^{\bar{i}}) = \frac{i}{2} \left( \begin{array}{cc}
            \sigma^{\bar{i}}  & 0 \\
            0 & \sigma^{\bar{i}} \\
          \end{array} \right),
 \quad
T_{10} = \frac{i}{2} (\sigma^2 \otimes \mathbf{I}_2) = \frac{1}{2} \left( \begin{array}{cc}
            0  & \mathbf{I}_2 \\
            - \mathbf{I}_2 & 0 \\
          \end{array} \right),
\eea
where $\sigma^{\bar{i}} \; ({\bar{i}} = 1, 2, 3)$ are the Pauli matrices.
The generators $T_a \; (a = 1, \cdots, 10)$ in \eq{lie-sp4} satisfy the commutation relation
\be \la{comm-sp4}
[T_a, T_b] = - {t_{ab}}^c T_c
\ee
with the totally antisymmetric structure constants (see Eq. (3.32) in \cite{psy-gist})
\be \la{str-cons-sp4}
t^{(2{\bar{i}}-1)(2{\bar{j}}-1){(6 + \bar{k})}} = t^{(2{\bar{i}})(2{\bar{j}}) {(6+\bar{k})}}
= t^{(6+{\bar{i}})(6+{\bar{j}}) {(6+\bar{k})}} = \varepsilon^{{\bar{i}}{\bar{j}}{\bar{k}}},
\quad  t^{(2\bar{i}-1)(2{\bar{j}}) 10} = \delta^{{\bar{i}}{\bar{j}}}.
\ee
We intentionally used a different symbol to distinguish the structure constants of $Sp(4)$
from those of $SU(4)$.
We take the Lie algebra generators in \eq{coset-comm} as follows:
\bea \la{sp4-cobasis}
 && \mathcal{H}_{\bar{a}} =  T_{\bar{a}}, \qquad  {\bar{a}} = (6 + {\bar{i}}, 10), \xx
 && \mathcal{K}_i =  (T_{2\bar{i}-1},  T_{2{\bar{j}}}),
 \qquad  i = (2{\bar{i}}-1, 2{\bar{j}}),
\eea
where ${\bar{i}}, {\bar{j}} = 1, 2, 3$.
The maximal subgroup $H = SU(2) \times U(1)$ is generated by
$T_{\bar{a}} =(T_{6+\bar{i}}, T_{10})$.
Thus the structure constants in \eq{coset-comm} are given by
\be \la{sp4-cos-con}
{C_{\bar{a} \bar{b}}}^{\bar{c}} = - t^{(6+{\bar{i}})(6+{\bar{j}}) {(6+\bar{k})}}, \quad
{C_{\bar{a} i}}^{j} = - t^{\bar{a} i j}, \quad
{C_{ij}}^{\bar{a}} = - t^{i j \bar{a}},
\ee
where $i, j = (2{\bar{i}}-1, 2{\bar{j}}) = 1, \cdots, 6$, and $t^{\bar{a} i j} = t^{i j \bar{a}}$
refer to the structure constants in \eq{str-cons-sp4}.

Define a two-form
\be \la{sp4-2form}
\Omega \equiv d \omega^{10} = \theta^1 \wedge \theta^2 + \theta^3 \wedge\theta^4 +  \theta^5 \wedge \theta^6,
\ee
where we used Eq. \eq{comp-cm} and the structure constants in \eq{str-cons-sp4}.
Note that $\Omega = d \omega^{10}$ provides a non-degenerate closed two-form,
i.e. $d \Omega = 0$, on $M$. Therefore we can use the two-form $\Omega$ to define
the HYM equations on $M$. We will show how it works.

Since ${C_{ij}}^{k} = 0$, the coset space $G/H$ is symmetric.
The connection one-form is then given by
\be \la{co-sp4-conn}
{\omega^i}_j  = t^{i j \bar{a}} \omega^{\bar{a}}.
\ee
Their components are listed in appendix B.2.
Using the results there, it is straightforward to identify $su(3)$ gauge fields
and their strengths. Only three components are non-vanishing:
\bea \la{sym-sp4-af}
&& A^{(-) \hat{2}} = A^{(-)14} = 2 \omega^9,
\quad A^{(-) \hat{5}} = - A^{(-)10} = 2 \omega^8,
\quad A^{(-) \hat{7}} = A^{(-)5} = - 2 \omega^7, \xx
&& F^{(-) \hat{2}} = F^{(-)14} = 2 (\theta^1 \wedge \theta^3 + \theta^2 \wedge \theta^4),
\quad F^{(-) \hat{5}} = - F^{(-)10} = - 2 (\theta^1 \wedge \theta^5 + \theta^2 \wedge \theta^6), \xx
&& F^{(-) \hat{7}} = F^{(-)5} = - 2 (\theta^3 \wedge \theta^5 + \theta^4 \wedge \theta^6).
\eea
The reason why $F^{(-) \hat{3}}$ and $F^{(-) \hat{8}}$ vanish is
because $A^{(-)3}$, $A^{(-)8}$, and $A^{(-)15}$ are basically $U(1)$ gauge fields
according to Eqs. \eq{cosp4-conn} and \eq{-sp4-curv},
so they cannot generate $su(3)$ gauge fields.
The list in Eq. \eq{sym-sp4-af} implies that the gauge fields are basically $SU(2)$ gauge fields
in $H = SU(2) \times U(1)$. It confirms that the natural $H$-structure on $G/H$
is contained within the $SU(3)$-structure \cite{lech-popo1}.
It is easy to check that the above connections satisfy the HYM
equations \eq{hym-su3} defined with the closed two-form \eq{sp4-2form}.
Therefore they describe a HYM instanton on $M = Sp(4)/[SU(2) \times U(1)]_{max}$
although $M$ is not a K\"ahler but nearly K\"ahler manifold.

\subsection{Non-symmetric $Sp(4)/[SU(2) \times U(1)]_{non-max}$}

We take the basis of the Lie algebra of $Sp(4)$ in Ref. \cite{mhs-coset}.
(See also Appendix A in \cite{dlust}.)
The structure constants in \eq{coset-comm} are given by
\bea \la{sp4-ns-str}
&& {C_{1,3}}^6 = - {C_{1,4}}^5 = {C_{2,3}}^5 = {C_{2,4}}^6 = 1, \xx
&& {C_{7,1}}^6 = - {C_{7,2}}^5 = {C_{8,1}}^5 = {C_{8,2}}^6 = {C_{9,1}}^2 = - {C_{9,5}}^6 =
{C_{10,1}}^2 = {C_{10,5}}^6 = 1, \xx
&&{C_{7,8}}^9 = {C_{10,3}}^4 = 2.
\eea
and they are totally antisymmetric but very different from the symmetric case \eq{str-cons-sp4}.
The indices $7, 8, 9$ and $10$ refer to the generators of the non-maximal subgroup
$H = SU(2) \times U(1)$. Note that the structure constants ${C_{ij}}^k$
(the first line in \eq{sp4-ns-str}) are non-vanishing, so $M = G/H$ is not symmetric.
In this case, we can easily incorporate the torsion by modifying the torsion-free
condition \eq{coset-tor} as \cite{dlust}
\be  \la{mod-tor}
T^i = d \theta^i + {\omega^i}_j \wedge \theta^j = \frac{\alpha}{2} {C^i}_{jk}
\theta^j \wedge \theta^k,
\ee
where $\alpha$ is an arbitrary real constant.
This modification corresponds to the replacement,
\be \la{torsion-rep}
{D_{kj}}^i \; \rightarrow \;
\frac{1}{2} (1 + \alpha) {C_{kj}}^i + {K_{kj}}^i,
\ee
in Eq. \eq{coeff-dk}. We will consider a general manifold with torsion.
The results for a six-manifold without torsion will be reproduced by setting $\alpha = 0$.
However, we will take a $G$-invariant metric as $g_{ij} = \delta_{ij}$ for simplicity.
In this case, the connection components ${K_{ij}}^k$ in \eq{coeff-dk}
identically vanish \cite{mhs-coset}, so the homogeneous Riemannian space $(G/H, g)$
is naturally reductive. After including the torsion, we can realize a Ricci-flat manifold
from the coset $M = G/H$ by choosing a specific parameter $\alpha$
if $M$ is a non-symmetric space with $\mathrm{rank} (G) = \mathrm{rank} (H)$ \cite{dlust}.

After including the torsion \eq{mod-tor}, the connection one-form is given by
\be \la{ns-sp4-conn}
{\omega^i}_j  = - C_{i j \bar{a}} \omega^{\bar{a}}
- \frac{1}{2} (1 + \alpha) C_{ij k} \theta^k.
\ee
Then the curvature two-forms are given by
\be \la{nc-sp4-curv}
R_{ij} = \frac{1}{2} \left( C_{ij \bar{a}}  C_{\bar{a} kl}
+ \frac{1}{2} (1 + \alpha) C_{ij m} C_{m kl}
+ \frac{1}{2} (1 + \alpha)^2 C_{ik m} C_{m l j} \right) \theta^k \wedge \theta^l.
\ee
The explicit results are presented in appendix B.2.3.
The Ricci tensor is determined by
\bea \la{ricci-nscp4}
R_{ik} &=& R_{ijkl} \delta^{jl} = C_{ij \bar{a}} C_{kj \bar{a}} + \frac{1}{4} (1 - \alpha^2) C_{ijm} C_{kjm} \xx
&=& (5 - \alpha^2 ) \delta^{ik}.
\eea
Therefore we see that the non-symmetric coset space $Sp(4)/[SU(2) \times U(1)]_{non-max}$
becomes Ricci-flat \cite{dlust} if we choose the torsion parameter $\alpha$ to be
\be \la{tor-alpha}
\alpha = \pm \sqrt{5}.
\ee

It is straightforward to calculate $SU(4)$ field strengths \eq{su4-field-id} using the $SU(4)$ gauge fields in \eq{ns-sp4-gauge}
and the result precisely coincides with the curvature tensors in \eq{nc-sp4-curv}
with the identification $F^{(\pm) a} = \eta^{(\pm)a}_{ij} R_{ij}$.
We can also identify the $U(1)$ gauge field defined by \eq{a0}. It is given by
\be \la{nssp4-a0}
A^{(\pm) 0} = - 4 \omega^{10}
\ee
and its field strength becomes
\be \la{nssp4-f0}
\widetilde{F}^{(\pm) 0} =  d A^{(\pm) 0} = 4 (\theta^1 \wedge \theta^2 + 2 \theta^3 \wedge \theta^4 + \theta^5 \wedge \theta^6 ).
\ee
This is a unique $Sp(4)$-invariant two-form that is closed, because the second Betti number is $b_2 (M) = 1$
on the underlying space $M$ \cite{mhs-coset}.
Note that it is different from the symmetric case \eq{sp4-2form}, especially, in the weights of 2-planes $([12], [34], [56])$.
This is due to the different weights in the $U(1)$ generator $T_{10}$ in $H = SU(2) \times U(1)$
as can be seen from the structure constants \eq{sp4-ns-str}.
This kind of invariant two-form also appears in another non-symmetric coset space $SU(3)/ U(1) \times U(1)$ \cite{mhs-coset}.

Some question naturally arises. Can we define HYM equations
using the invariant two-form $\widetilde{\Omega} \equiv \theta^1 \wedge \theta^2 + 2 \theta^3 \wedge \theta^4
+ \theta^5 \wedge \theta^6$ in \eq{nssp4-f0}?
In other words, can we define the Bogomolny equation \eq{bogomolny} with the two-form $\widetilde{\Omega}$?
We have not been able to derive such a consistent equation with the two-form $\widetilde{\Omega}$.
In particular, it turns out that it is difficult to satisfy
Eq. \eq{stable-eq} with $\widetilde{\Omega}$, which is known as the stability condition of vector bundles.
But, the two-form $\widetilde{\Omega}$ may be allowed if a six-dimensional manifold is a
direct product of a four-dimensional manifold and a two-dimensional Riemann surface;
$M_6 = M_4 \times \Sigma_2$ \cite{cdfn,bakasi}.
Thus we conclude that the HYM equations \eq{hym-eq}
can only be defined with the canonical two-form $\Omega$.

The $SU(4)$ field strengths displayed in \eq{nssp4-field-comp} exhibit an interesting property.
The members in $\Omega^2_{\pm 6}$, i.e. $F^{(+)1,2,4,5,9,10}$, satisfy the $\Omega$-self-duality equations,
$F^{(+)i} = *(F^{(+)i} \wedge \Omega)$, while the members in $\Omega^2_{\pm 8}$, i.e. $F^{(+)6,7,11,12,13,14}$,
satisfy the $\Omega$-anti-self-duality equations, $F^{(+)\hat{a}} = - *(F^{(+)\hat{a}} \wedge \Omega)$,
except $F^{(+)3,8,15}$. But the members in $\Omega^2_{\pm 8}$ for
$F^{(\pm)3,8,15}$ (see \eq{su3-field} for their definition) can be written as
\bea \la{asd-38}
&& F^{(\pm)\hat{3}} = -2 (2[12] - [34] - [56]) + \frac{1}{2} (1+\alpha)^2 ([12]+[34]-2[56])
+ (1-\alpha)(3+\alpha) [12], \xx
&& F^{(\pm)\hat{8}} = - \frac{2}{\sqrt{3}} ([12] + [34] - 2 [56]) + \frac{1}{2\sqrt{3}} (1+\alpha)^2 (2[12] - [34] - [56]) \xx
&& \hspace{1.5cm} - \frac{1}{2\sqrt{3}} (1-\alpha)(3+\alpha) ([12] + [56]),
\eea
where we used a shortand notation $[ij] : = \theta^i \wedge \theta^j$.
One can see that $F^{(\pm)\hat{3}}$ and $F^{(\pm)\hat{8}}$ also satisfy the $\Omega$-anti-self-duality equations,
$F^{(\pm)\hat{a}} = - *(F^{(\pm)\hat{a}} \wedge \Omega)$,
for $\alpha = 1$ or $\alpha = -3$. These values are different from the value
\eq{tor-alpha} for the Ricci-flatness. Therefore the field strengths $F^{(\pm)\hat{a}}$
for $\alpha = 1, -3$ describe a HYM instanton.
The meaning of $F^{(\pm)i}$'s obeying the $\Omega$-self-duality equations is not obvious
but they may correspond to deformations of the HYM instanton.
It is very interesting that there exists no HYM instanton
on the coset space $Sp(4)/[SU(2) \times U(1)]_{non-max}$ without introducing the torsion.

There is an intriguing property for $U(1)$ gauge fields.
The $U(1)$ gauge fields defined by \eq{a0} have the field strength
given by
\be \la{intrig-u1}
F^{(\pm)0} = 4([12] + 2 [34] + [56]) - (1 +\alpha)^2 [34]
= 4 \widetilde{\Omega} - (1 +\alpha)^2 [34]
\ee
which can be computed by using the result \eq{nssp4-field-comp}.
It is different from \eq{nssp4-f0} except $\alpha = -1$, in which
the connection \eq{ns-sp4-conn} becomes an $H$-connection.
Furthermore, HYM instantons exist only for $\alpha = 1$
or $\alpha = -3$ where the field strength \eq{intrig-u1} becomes $F^{(\pm)0} = 4 \Omega$.
This is the reason why we use a different notation in \eq{nssp4-f0}.

\section{Topological invariants}

The Euler characteristic $\chi(M)$ for a six-dimensional Riemannian manifold $M$ is given by \cite{nakahara}
\be \la{euler-number}
\chi(M) = \frac{1}{(4\pi)^3 3!} \int_M \varepsilon^{A_1 A_2 \cdots A_6} R_{A_1 A_2}
\wedge R_{A_3 A_4} \wedge R_{A_5 A_6}.
\ee
Using the results \eq{cp3-curv} and \eq{sp4-curv}, one can explicitly calculate the Euler characteristic
\eq{euler-number} for $\mathbb{C}\mathbb{P}^3$. The result is $\chi(M) = 4$ for both cases as expected. But the details are different:
\bea \la{euler-vol}
&& \rho_\chi = \left\{
  \begin{array}{ll}
    2^{10} \times 3^2 \times \mu(M), & \hbox{$M = SU(4)/U(3)$;} \\
    2^{6} \times 3^2 \times \mu(M), & \hbox{$M = Sp(4)/[SU(2) \times U(1)]_{max}$,}
  \end{array}
\right. \xx
&& \mathrm{vol}(M) = \int_M \mu(M) = \left\{
  \begin{array}{ll}
    \frac{\pi^3}{6}, & \hbox{$M = SU(4)/U(3)$;} \\
     \frac{8 \pi^3}{3}, & \hbox{$M = Sp(4)/[SU(2) \times U(1)]_{max}$,}
  \end{array}
\right.
\eea
where $\rho_\chi \equiv \varepsilon^{A_1 A_2 \cdots A_6} R_{A_1 A_2}
\wedge R_{A_3 A_4} \wedge R_{A_5 A_6}$ and $\mu(M) \equiv \theta^1 \wedge \cdots \wedge \theta^6$ is a volume form.
For the first case $M = SU(4)/U(3)$, 15 terms contribute in total where the single term $R_{12} \wedge R_{34} \wedge R_{56}$
contributes $2/3$ and remaining 14 terms contribute $1/3$ to $\chi(M)$.
On the other hand, for the second case $M = Sp(4)/[SU(2) \times U(1)]_{max}$,
only 4 terms contribute where the single term $R_{12} \wedge R_{34} \wedge R_{56}$
contributes one half and the remaining 3 terms contribute the other half to $\chi(M)$.

The non-symmetric coset space in Sect. 5.3 is particularly interesting because we have included the torsion \eq{mod-tor}
in the connection. Now every 15 terms contribute to $\chi(M)$ where each term depends on the torsion parameter $\alpha$,
but adding them together leads to $\rho_\chi = 2^{8} \times 3^2 \times \mu(M)$, completely independent of $\alpha$.
Then the integral \eq{euler-number} gives us the number $\chi(M) = 4$ where we used the volume formula,
$\mathrm{vol}(M) = \frac{2}{3} \pi^3$ \cite{mhs-coset}.
Despite the fact that the topological number $\chi(M)$ cannot depend on a continuous parameter
such as $\alpha$, the (magic) cancellation of the torsion parameter in $\rho_\chi$ is quite interesting.

A solution of the HYM equations \eq{hym-su3} carries two topological numbers
in the second and third Chern classes \cite{gsw-vol2}.
The first invariant is the instanton number defined by the second Chern class of $G$-bundle $E$ as \cite{gang-tian}
\be \la{2-chern}
C_2(E) \cdot [\Omega] = \frac{1}{8 \pi^2} \int_M \Tr F^{(\pm)} \wedge F^{(\pm)} \wedge \Omega
= - \frac{1}{16 \pi^2} \int_M F^{(\pm)a} \wedge F^{(\pm)a} \wedge \Omega,
\ee
where $[\Omega]$ denotes the cohomology class of $\Omega$.
The second one is the Euler number defined by the third Chern class
of $G$-bundle $E$ as \cite{egh-report,nakahara}
\be \la{3-chern}
C_3(E) = \frac{i}{24 \pi^3} \int_M \Tr F^{(\pm)} \wedge F^{(\pm)} \wedge F^{(\pm)}
= \pm \frac{1}{96 \pi^3} \int_M d^{abc} F^{(\pm)a} \wedge F^{(\pm)b} \wedge F^{(\pm)c}.
\ee
The third Chern number $C_3(E)$ is equal to the Euler number \eq{euler-number}
if the vector bundle $E$ is a tangent bundle on $M$, i.e. $E = TM$.
One can see this identity by using $R_{AB} = F^{(\pm)a} \eta^{(\pm)a}_{AB}$ and
$\frac{1}{4} \varepsilon^{A_1 A_2 \cdots A_6} \eta^{(\pm)a}_{A_1 A_2}
\eta^{(\pm)b}_{A_3 A_4} \eta^{(\pm)c}_{A_5 A_6} = d^{abc}$. See
(B.9) and (B.10) in \cite{yang-yun}. As we mentioned above, this number is 4
for all cases we have considered. Using the results in appendix B,
we can also calculate the instanton number \eq{2-chern}:
\be \la{inst-den}
\rho_I = \left\{
  \begin{array}{lll}
    192 \mu(M), & \hbox{$M = SU(4)/U(3)$;} \\
    12 \mu(M), & \hbox{$M = Sp(4)/[SU(2) \times U(1)]_{max}$;} \\
    96 \mu(M) - 12 (1+\alpha)^2 \mu(M), & \hbox{$M = Sp(4)/[SU(2) \times U(1)]_{non-max}$,}
  \end{array}
\right.
\ee
where $\rho_I \equiv F^{(\pm)a} \wedge F^{(\pm)a} \wedge \Omega$. Then we get
$C_2(E) \cdot [\Omega] = - 2\pi$ for the first two cases, while
$C_2(E) \cdot [\Omega] = - 4\pi + \frac{\pi}{2} (1 + \alpha)^2$ for the last case.
The instanton number appears to depend on the torsion parameter $\alpha$.
This is probably because the structure of the vector bundle $E$ depends on
the torsion parameter. For example, an instanton solution exists only for certain
values of $\alpha$ as was shown before.
But, for the vector bundle $E$ with $\alpha = 1$
or $\alpha = -3$ which admits a HYM instanton, it also reduces to
$C_2(E) \cdot [\Omega] = - 2\pi$. For all cases, the instanton number is equal
to $C_2(E) = - 1$ if the cohomology class of $\Omega$ is normalized to $\left[ \frac{\Omega}{2\pi} \right] = 1$.

Now, let us calculate the instanton number $C_2(E)$ and the Euler number $C_3(E)$ for
the HYM instantons.
It is known \cite{gibb-pope,oh-yang} that the instanton number of the Yang-Mills
solution derived from a gravitational metric is not an integer.
For $\mathbb{C}\mathbb{P}^2$, for instance, $C_2(E_+) = \frac{9}{4}$ and $C_2(E_-) = -\frac{3}{4}$,
where $E_\pm$ are (anti-)self-dual instanton bundles and $E_+$ is essentially a $U(1)$-bundle.
The Euler characteristic $\chi(M)$ for a four-manifold $M$
is generally given by $\chi(M) = C_2(E_+) - C_2(E_-)$, e.g.
$\chi(\mathbb{C}\mathbb{P}^2) = \frac{9}{4} + \frac{3}{4} = 3$ \cite{oh-yang}.
This pathology was interpreted in \cite{gibb-pope} as a local nature of the solution
so that it is not globally well-defined. Notice that the topological
numbers defined in \eq{2-chern} and \eq{3-chern} consistently take integer values
for $SU(4)$ connections. However, if we separate $SU(3)$ connections from the $SU(4)$ connections
in order to construct HYM instantons, the topological numbers
for these instanton connections will generally not take integer values.
Therefore, it is more reasonable to interpret the fractional topological charge
as a part of the contribution of $SU(3)$ instanton connections to the total topological charge.
The contributions from each part need not be quantized into integers,
but it is enough for the total contribution of all parts to be quantized to an integer.
This interpretation also applies to the four-dimensional case,
and the sum in $\chi(\mathbb{C}\mathbb{P}^2) = \frac{9}{4} + \frac{3}{4} = 3$ reflects this interpretation precisely.
We will verify this interpretation for the six-dimensional case.

It is straightforward to calculate the instanton number $C_2(E)$ and the Euler number $C_3(E)$
for the HYM instantons using Eqs. \eq{su3-field},
\eq{sym-sp4-af}, \eq{nssp4-field-comp} and \eq{asd-38}:
\bea \la{inst-hym}
C_2(HYM) \cdot [\Omega]
&=& - \frac{1}{16 \pi^2} \int_M F^{(\pm)\hat{a}} \wedge F^{(\pm)\hat{a}} \wedge \Omega  \xx
&=& \left\{
  \begin{array}{lll}
    \frac{2}{3} \pi, & \hbox{$M = SU(4)/U(3)$;} \\
    4 \pi, & \hbox{$M = Sp(4)/[SU(2) \times U(1)]_{max}$;} \\
    \frac{8}{3} \pi, & \hbox{$M = Sp(4)/[SU(2) \times U(1)]_{non-max}$,}
  \end{array}
\right. \\
\la{euler-hym}
C_3(HYM) &=& \pm \frac{1}{96 \pi^3} \int_M d^{\hat{a}\hat{b}\hat{c}} F^{(\pm)\hat{a}} \wedge F^{(\pm)\hat{b}} \wedge F^{(\pm)\hat{c}} \xx
&=& \left\{
  \begin{array}{lll}
    \pm \frac{20}{27}, & \hbox{$M = SU(4)/U(3)$;} \\
    0, & \hbox{$M = Sp(4)/[SU(2) \times U(1)]_{max}$;} \\
    \pm \frac{28}{27}, & \hbox{$M = Sp(4)/[SU(2) \times U(1)]_{non-max}$,}
  \end{array}
\right.
\eea
where $\hat{a}, \, \hat{b}, \, \hat{c} = 1, \cdots, 8$ are $su(3)$ Lie algebra indices
and the last case has been calculated with the torsion parameter $\alpha = 1, -3$.
Note that the third Chern number $C_3(HYM)$ for the symmetric coset space $M = Sp(4)/[SU(2) \times U(1)]_{max}$ vanishes.
This is due to the fact that Eq. \eq{sym-sp4-af} basically describes $SU(2)$ gauge fields
and the structure constant $d^{\hat{a}\hat{b}\hat{c}}$ for the $SU(2)$ group identically vanishes.

In order to check our claim above, we also compute the contributions
to the topological numbers from the complement of $su(3) \subset su(4)$.
We have identified the contribution from the complement using \eq{exp-3chern}.
Using the result and $d^{0 0 \hat{a}} = 0$, we get
\bea \la{inst-u1}
C_2 (U(1)) \cdot [\Omega]
&=& - \frac{1}{16 \pi^2} \left( \frac{2}{3} \right) \int_M F^{(\pm) 0} \wedge F^{(\pm)0} \wedge \Omega  \xx
&=& \left\{
  \begin{array}{ll}
    - \frac{8}{3} \pi,  & \hbox{$M = SU(4)/U(3)$;} \\
    - 6 \pi, & \hbox{$M = Sp(4)/[SU(2) \times U(1)]_{max}$,}
  \end{array}
\right. \\
\la{euler-u1}
C^\perp_3 (E) &=& \pm \frac{1}{96 \pi^3} \int_M \left( \left(\frac{2}{3} \right)^{\frac{3}{2}} d^{000} F^{(\pm)0} \wedge F^{(\pm)0} \wedge F^{(\pm)0}
+ \sqrt{6} d^{0 \hat{a} \hat{b}} F^{(\pm)0} \wedge F^{(\pm) \hat{a}} \wedge F^{(\pm)\hat{b}} \right) \xx
&=& \pm \frac{1}{96 \pi^3} \int_M \left( \frac{4}{9} F^{(\pm)0} \wedge F^{(\pm)0} \wedge F^{(\pm)0}
- F^{(\pm)0} \wedge F^{(\pm) \hat{a}} \wedge F^{(\pm)\hat{a}} \right) \xx
&=& \left\{
  \begin{array}{lll}
    \pm \frac{(64 + 24)}{27}, & \hbox{$M = SU(4)/U(3)$;} \\
    \pm (2+2), & \hbox{$M = Sp(4)/[SU(2) \times U(1)]_{max}$.}
  \end{array}
\right.
\eea
The normalization factor $2/3$ in \eq{inst-u1} for the $U(1)$ generator
can be fixed by noting that
$$ F^{(\pm)3} \wedge F^{(\pm)3} + F^{(\pm)8} \wedge F^{(\pm)8}
+  F^{(\pm)15} \wedge F^{(\pm)15} = \frac{2}{3} F^{(\pm)0} \wedge F^{(\pm)0} +  F^{(\pm)\hat{3}} \wedge F^{(\pm)\hat{3}}
+ F^{(\pm)\hat{8}} \wedge  F^{(\pm) \hat{8}}. $$
In order to calculate $C^\perp_3 (E)$ in \eq{euler-u1}, we have used
the structure constants in \eq{st-con-0}.
Then the second term in \eq{euler-u1} is proportional to the second Chern number of the HYM instanton in \eq{inst-hym}.
One can immediately see that the sum of the contributions from the two parts,
i.e. \eq{inst-hym} + \eq{inst-u1} and \eq{euler-hym} + \eq{euler-u1}, correctly reproduce the topological charges for the $SU(4)$ connections.
In particular, the Euler number of the coset space $Sp(4)/[SU(2) \times U(1)]_{max}$ completely comes
from the third Chern class of $U(1)$ gauge fields and the second Chern class of the $SU(2)$ Yang-Mills instanton.

For the non-symmetric coset space $M = Sp(4)/[SU(2) \times U(1)]_{non-max}$,
the complement of $su(3) \subset su(4)$ contains the dotted part
in $su(4) - u(3)$ as well as the $U(1)$ part.
Thus we need to calculate the contribution from the components $F^{(+)1,2,4,5,9,10}$
in $\Omega^2_{\pm 6}$ that satisfy the $\Omega$-self-duality equations, $F^{(+)i} = *(F^{(+)i} \wedge \Omega)$,
as well as the $U(1)$ component $F^{(+)0} = 4 \Omega$. The result is
\bea \la{topnum-nscp3}
&& C^\perp_2 (E) \cdot [\Omega] = - \frac{14}{3} \pi, \xx
&& C^\perp_3 (E) = \pm \frac{80}{27}.
\eea
Interestingly, the dotted part also contributes $\frac{3}{7}$ to $C^\perp_2 (E)$ compared to the $U(1)$ part.
In order to calculate $C^\perp_3 (E)$ in \eq{exp-3chern}, we have used the structure constants
\be \la{ns-stc-d}
d^{00 \dot{a}} = d^{0 \hat{a} \dot{a}} =  d^{0 \dot{a} \hat{a}} = d^{\hat{a} \hat{b} \dot{a}} = d^{\dot{a} \dot{b} \dot{c}} = 0,
\quad d^{0 \dot{a} \dot{b}} = \frac{1}{\sqrt{6}} \delta^{\dot{a} \dot{b}}
\ee
in addition to \eq{st-con-0}.
It turns out that the contribution from the $d^{0 \dot{a} \dot{b}}$-part exactly cancels out
the contribution from the $d^{\hat{a} \dot{a} \dot{b}}$-part. Consequently, the dotted space does not contribute
to $C_3 (E)$ at all unlike the second Chern number $C_2 (E)$.
Again the sum of the contributions from all parts exactly recovers
the total integer charges.

\section{Discussion}

Compactifying a heterotic string theory from ten to four dimensions imposes several critical constraints
to ensure consistency and phenomenological viability \cite{gsw-vol2}.
The ten-dimensional heterotic string theory has an anomaly cancellation mechanism
involving the gauge group and the spin connection characterized by the subtle identity \eq{anomaly-string}.
The condition \eq{anomaly-string} enforces that the cohomology class of the first Pontryagin class of
the tangent bundle $TM$ must belong to the same cohomology class of the second Chern class
of the principal $G$-bundle where $G$ is $SO(32)$ or $E_8 \times E_8$.
For a six-dimensional manifold $M$, the spin connection is an $SO(6)\cong SU(4)/\mathbb{Z}_2$
or $Spin(6) \cong SU(4)$ gauge field.
Thus, in order to satisfy the constraint \eq{anomaly-string}, it is necessary to identify the $SU(4)$ gauge field in $TM$
with the $G$-gauge field. For Calabi-Yau manifolds, for example, the embedding $SU(3) \hookrightarrow G$ is necessary.
The generalized 't Hooft symbols \eq{int-tsymbol} provide a compact and efficient way to identify $SU(3)$ connections
from $SO(6)$ spin connections. Furthermore, the six-dimensional self-dual structures governed by the HYM equations
can be elegantly organized using them.

In four dimensions, the 't Hooft symbols $\eta^{(\pm)a}_{AB} \; (a=1,2,3; A, B = 1, \cdots,4)$
are either self-dual or  anti-self-dual, i.e.
$\eta^{(\pm)a}_{AB} = \pm \frac{1}{2} \varepsilon^{ABCD} \eta^{(\pm)a}_{CD}$.
So they provide canonical bases of (anti-)self-dual two-forms, $\zeta^a_\pm = \frac{1}{2} \eta^{(\pm)a}_{AB} e^A \wedge e^B$,
obeying the Hodge-duality $ * \zeta^a_\pm = \pm \zeta^a_\pm$. On an orientable four-manifold,
the vector space of two-forms then decomposes into the spaces of self-dual and anti-self-dual two-forms:
\be \la{4d-2form}
\Omega^2 (M) = \Omega^2_+ (M)  \oplus \Omega^2_- (M),
\ee
where $\Omega^2_\pm (M)$ are the $\pm 1$ eigenspaces of the Hodge star operator $*: \Omega^2 (M) \to \Omega^2 (M)$.
A four-dimensional spinor representation of $Spin(4) = SU(2)_+ \times SU(2)_-$ also has two irreducible
spinor reprsentations $S_\pm$, each of dimension 2,
whose Lie algebra generators are given by $2 \times 2$ matrices $J_\pm^{AB}$.
According to the product structure of the Lorentz group $Spin(4) = SU(2)_+ \times SU(2)_-$,
the vector space of $so(4)$ Lie algebra splits into the same structure as the two-form vector space \eq{4d-2form}:
\be \la{4d-lie}
so(4) \cong spin(4) = su(2)_+ \oplus su(2)_-.
\ee
Since $J_\pm^{AB}$ are traceless anti-Hermitian matrices, they can be expanded in the basis of Pauli matrices:
\be \la{4d-thooft}
J_\pm^{AB} = \frac{i}{2} \eta^{(\pm)a}_{AB} \sigma^a.
\ee
Then the 't Hooft symbols are determined by $\eta^{(\pm)a}_{AB} = -i \Tr (J_\pm^{AB} \sigma^a)$.
These 't Hooft symbols have been extensively used in \cite{yang-gist,yang-coll23} to study the algebraic
and topological properties of four-dimensional Riemannian manifolds.

The 't Hooft symbols in four and six dimensions represent the linear transformation between the Lie algebra generators of an irreducible representation of the Lorentz group and those of its isomorphic classical group.
This irreducible representation plays a crucial role in determining the structure of the 't Hooft symbols.
But there are subtle differences between four and six dimensions.
Since the Lie algebra generator of an irreducible spinor representation in six dimensions takes the form
$J_\pm^{AB} \sim \Gamma^{AB} \pm \frac{1}{4!} \varepsilon^{ABCDEF} \Gamma^{CDEF}$,
the vector space isomorphism implies that the corresponding irreducible representation
for the vector space of two-forms takes the following split \cite{yang-yun}
\be \la{6d-2dec}
\Omega^2 (M) \oplus * \Omega^4 (M)  = \Omega^2_+ (M)  \oplus \Omega^2_- (M).
\ee
The canonical basis of the irreducible vector space $\Omega^2_\pm (M)$ is given by \eq{2-forms}.
Each vector space $\Omega^2_\pm (M)$ of dimension 15 is further decomposed into the eigenspaces
of the operator \eq{w-hodge} of dimension 1, 6 and 8, as shown in \eq{2form-decomp}.

What is the meaning of the splitting \eq{6d-2dec}? The splitting in \eq{4d-2form} corresponds
to the eigenspaces of instantons and anti-instantons in four dimensions.
However, the splitting in \eq{6d-2dec} does not correspond to the splitting of instantons and anti-instantons.
As was shown here, each space $\Omega^2_\pm (M)$ contains a solution of instantons.
What distinguishes the two is not the instanton but the third Chern number.
This may be seen from the formula \eq{vol-id}.
For Calabi-Yau manifolds, Eq. \eq{mirror-euler} shows that a mirror pair have an opposite sign for the Euler number.
This is another reason why the mirror structure of Calabi-Yau manifolds may be captured
by two kinds of 't Hooft symbols defined in the chiral and anti-chiral sectors.

We showed that a solution of HYM instantons can be embedded in either $\Omega^2_+ (M)$ or $\Omega^2_- (M)$.
In this paper, we have considered the embedding of the same solution into two vector spaces, independently,
and confirmed in section 6 that they have the opposite third Chern number.
Of course, the sign flip in this case is trivial because complex conjugation
reverses the chirality (or vice versa) and it corresponds to an orientation reversing.
But we can embed two different solutions of HYM instantons into $\Omega^2_+ (M)$ and $\Omega^2_- (M)$, separately.
If they satisfy the relation \eq{mirror-euler}, they are mirror each other.
Although the splitting structure in \eq{6d-2dec} does not explain why (at least) Calabi-Yau manifolds
always exist with a mirror pair, it guarantees that it is always possible to accommodate a mirror manifold
with the splitting structure. We will address this issue in a follow-up paper.

The anatomy of topological invariants shows many interesting aspects.
First we observe that even though the total instanton number is negative,
the second Chern number from HYM instantons is always positive as can be seen from \eq{inst-hym}
whereas the contribution from the $U(1)$ part (and dotted part) is negative.
However, we observe that all parts contribute with the same sign to the third Chern number.
In particular, for the non-symmetric coset space $M = Sp(4)/[SU(2) \times U(1)]_{non-max}$,
the dotted space has no contribution to the third Chern number although it is nontrivial.
Moreover the third Chern number does not depend on the torsion parameter $\alpha$.
Note that the spin connection \eq{ns-sp4-conn} becomes an $H$-connection for $\alpha = -1$
and Eq. \eq{nssp4-field-comp} shows that nonvanishing field strengths are only
$F^{(+)3,6,7,8}$ or $F^{(-)3,6,7,8, 15}$. But they are not a HYM instanton
since they do not satisfy \eq{hym-eq}.
We can test why the dotted space does not contribute to the third Chern number \eq{3-chern}
by computing it for $\alpha = -1$.
It is easy to check that the number is 4 as expected.

As we observed above, the instanton number from HYM instantons is always positive.
This fact will be important for our next studies.
Suppose that we consider a ten-dimensional Einstein-Yang-Mills theory.
Consider a time-dependent warped metric of the form
\be \la{10-inflation}
ds^2 = - dt^2 + a(t)^2 d \mathbf{x} \cdot d \mathbf{x} + b(t)^2 ds_6^2 (\mathbb{C}\mathbb{P}^3),
\ee
where $a(t)$ and $b(t)$ are arbitrary time-dependent functions and
$ds_6^2 (\mathbb{C}\mathbb{P}^3)$ is, for example, the Fubini-Study metric \eq{cp3-metric}.
If we put a HYM instanton on $\mathbb{C}\mathbb{P}^3$, only the last topological term in the Yang-Mills action \eq{bogomolny}
will not vanish. And this topological term will act as a cosmological constant in four-dimensional spacetime
and trigger a cosmic inflation of our universe.
We guess that this model will also show a similar behavior as the eight-dimensional model \cite{kky,hkky}
so that the  cosmic inflation of the four-dimensional universe simultaneously
causes a dynamical compactification of the extra dimensions.
If this ten-dimensional inflation model works successfully, we think we can incorporate this model
into a ten-dimensional (heterotic) string theory. This issue will also be addressed in our next project.

\section*{Acknowledgments}

We thank Jun Soo Seol and Jaewon Shin for their significant contribution at the early stage of this work.
We are also deeply grateful to Prof. Youngjoo Chung for his numerical check using MathSymbolica.
This work was supported by the National Research Foundation of Korea (NRF) with grant number NRF-2018R1D1A1B0705011314.

\appendix

\section{Six-dimensional 't Hooft symbols}

The Lorentz generators in $spin(6)$ Lie algebra are given by
\begin{equation}\label{lorentz-gen}
J^{AB} = \frac{1}{4}[\Gamma^A, \Gamma^B],
\end{equation}
where $\Gamma^A \; (A =1, \cdots, 6)$ are Gamma-matrices obeying the Dirac algebra
\begin{equation}\label{dirac-alg}
\{\Gamma^A, \Gamma^B \} = 2 \delta^{AB} \textbf{I}_8.
\end{equation}
In six dimensions, the spinor representation is reducible and its irreducible representations are given by positive and negative chiral representations.
This splitting is due to the existence of a nontrivial chiral operator
$\Gamma_7 \equiv i \Gamma^1 \cdots \Gamma^6$ with the property $\Gamma_7^2 = \textbf{I}_8$. The projection operators defined by
\begin{equation} \label{gamma7}
\Gamma_\pm = \frac{1}{2} (\textbf{I}_8 \pm \Gamma_7)
\end{equation}
split the Lorentz generators in \eq{lorentz-gen} into chiral generators
\begin{equation} \label{chiral-gen}
J^{AB}_\pm \equiv \Gamma_\pm J^{AB}, \qquad \Gamma_7 J^{AB}_\pm = \pm J^{AB}_\pm.
\end{equation}
The chiral generators separately satisfy the Lorentz algebras:
\begin{eqnarray}
 && [J^{AB}_\pm, J^{CD}_\pm]  = - \left( \delta^{AC} J^{BD}_\pm - \delta^{AD} J^{BC}_\pm
 - \delta^{BC} J^{AD}_\pm + \delta^{BD} J^{AC}_\pm \right), \xx
 && [J^{AB}_\pm, J^{CD}_\mp]  = 0.
\end{eqnarray}

There exists an isomorphism \eq{spin6=su4} between the six-dimensional Lorentz group and
a classical Lie group.
According to this isomorphism, $spin(6)$ Lorentz algebra is isomorphic to $su(4)$ Lie algebra.
The splitting in \eq{chiral-gen} implies that $su(4)$ Lie algebra has a similar splitting.
Indeed, the positive $(S_+)$ and negative $(S_-)$ chiral spinors of $spin(6)$ can be identified
with the fundamental $(\textbf{4})$ and anti-fundamental $(\overline{\textbf{4}})$ representations of $su(4)$, respectively.
Note that the fundamental representation $\textbf{4}$ of $SU(4)$ is a complex representation
and its complex conjugate $\overline{\textbf{4}}$ is an inequivalent  representation different from $\textbf{4}$.
Therefore, we identify their Lie algebra generators as follows:
\begin{equation} \label{iden-liealg}
  J^{AB}_+ \cong T^a_+, \qquad J^{AB}_- \cong T^a_-,
\end{equation}
where $T^a_+ = \frac{i}{2} \lambda_ a \; (a = 1, \cdots, 15)$ are $su(4)$ Lie algebra generators
and $T^a_- =(T^a_+)^*$ and they obey the commutation relations
\begin{equation} \label{comm-su4}
[T^a_\pm, T^b_\pm] = - f^{abc} T^c_\pm.
\end{equation}
Here we follow Ref. \cite{yang-yun} for the expressions of $spin(6)$ and $su(4)$ Lie algebra generators.
The Lie algebra generators $(J^{AB}_\pm, T^a_\pm)$ constitute a complete basis
of $4 \times 4$ anti-Hermitian, traceless matrices. Thus the chiral generators $J^{AB}_\pm$
can be expanded in the basis of the $su(4)$ Lie algebra generators $T^a_\pm$
\begin{equation} \label{lin-exp}
J^{AB}_\pm = 2 \eta^{(\pm)a}_{AB} T^a_\pm,
\end{equation}
where the expansion coefficients $\eta^{(\pm)a}_{AB}$ are determined by \eq{int-tsymbol}.

This six-dimensional version of the 't Hooft symbols was first introduced in \cite{yang-yun}.
The algebraic relations obeyed by the 't Hooft symbols have been derived in appendix B
in \cite{yang-yun}. They provide an antisymmetric 6-dimensional representation of $SU(4)$
whose Lie algebra generators are defined by \eq{6dim-su4}.
They also provide a 15-dimensional basis of two-forms in $\Omega^2(M) = \Lambda^2 T^* M$
for each chirality:
\begin{equation} \label{2-forms}
J^a_\pm = \frac{1}{2} \eta^{(\pm)a}_{AB} e^A \wedge e^B.
\end{equation}
And they satisfy the useful identity
\begin{equation} \label{vol-id}
J_\pm^a \wedge J_\pm^b \wedge J_\pm^c = \pm \frac{1}{2} d^{abc} \mathrm{vol} (g)
\end{equation}
where $\mathrm{vol} (g) = \sqrt{g} d^6 x$ and $d^{abc} = \pm 4 i \Tr \left( T^a_\pm T^b_\pm T^c_\pm \right) $
are symmetric structure constants of $SU(4)$.

There is a unique two-form of rank 6
\begin{equation}
    \Omega = \frac{1}{2} I_{AB} e^A \wedge e^B = e^1 \wedge e^2 + e^3 \wedge e^4 + e^5 \wedge e^6,
\end{equation}
where $I_{AB} = (\textbf{I}_3 \otimes i \sigma^2)_{AB} =
2 \left( \pm \eta^{(\pm)3}_{AB} + \frac{1}{\sqrt{3}} \eta^{(\pm)8}_{AB}
+ \frac{1}{\sqrt{6}} \eta^{(\pm)15}_{AB} \right)$.
The remaining 14 two-forms can be classified by noting the structure of the 't Hooft symbols $\eta^{(\pm)a}_{AB}$.
They can be classified into two types according to the structure of their tensor products (see appendix B in \cite{yang-yun} for the explicit
representation of $\eta^{(\pm)a}_{AB}$):
\bea \la{eta-tstr}
&& A: \eta^{(\pm)a}_{AB} \sim \{ \lambda^{\hat{a}} \otimes \mathbf{I}_2, \; \lambda^{\hat{a}} \otimes \sigma^2 \}, \xx
&& B: \eta^{(\pm)a}_{AB} \sim \{ \lambda^{\hat{a}} \otimes \sigma^1,
\; \lambda^{\hat{a}} \otimes \sigma^3 \},
\eea
where $\lambda^{\hat{a}} \;(\hat{a} = 1, \cdots, 8)$ are $SU(3)$ Gell-Mann matrices and
$\sigma^i \; (i=1,2,3)$ are Pauli matrices.
Since $I = \textbf{I}_3 \otimes i \sigma^2$, they satisfy the commutation relations
\bea \label{icoset}
&& A: [I, \mathcal{T}^{(\pm)a}] = 0, \\
\label{iu3}
&& B: \{I, \mathcal{T}^{(\pm)a}\}  = 0,
\eea
where $\mathcal{T}^{(\pm)a}$ are the $su(4)$ generators \eq{6dim-su4} defined by the 't Hooft symbols.
One can show (see Eqs. (3.77) and (3.81) in \cite{yang-yun}) that, if $\mathcal{T}^{(\pm)a} \in B$,
they belong to the coset $\mathbb{C}\mathbb{P}^3 = SU(4)/U(3)$, while the type $A$ satisfies
$[I, \mathcal{T}^{(\pm)a}] = 0, \; a \in u(3) \subset su(4)$.
This property was also noticed in \cite{aru-fro} (see the appendix A).
Consequently, on any six-dimensional manifold $M$, the space of two-forms $\Omega^2_\pm(M)$ can be decomposed into three subspaces
\be \la{decom-2form}
\Omega^2_\pm(M) = \Omega^2_{\pm 1} \oplus \Omega^2_{\pm 6} \oplus \Omega^2_{\pm 8}.
\ee
The spaces in \eq{decom-2form} are locally spanned by
\bea \label{168-decom}
&& \Omega^2_{\pm 1} =  \Omega = e^1 \wedge e^2 + e^3 \wedge e^4 + e^5 \wedge e^6, \xx
&& \Omega^2_{+ 6} = \{J_+^1, J^2_+, J^4_+, J^5_+, J^9_+, J^{10}_+ \}, \xx
&& \Omega^2_{- 6} = \{J_-^1, J^2_-, J^6_-, J^7_-, J^{11}_-, J^{12}_- \}, \\
&& \Omega^2_{+ 8} = \{J_+^6, J^7_+, J^{11}_+, J^{12}_+, J^{13}_+, J^{14}_+, K_+, L_+ \}, \xx
&& \Omega^2_{- 8} = \{J_-^4, J^5_-, J^{9}_-, J^{10}_-, J^{13}_-, J^{14}_-, K_-, L_- \}, \nonumber
\eea
where $K_\pm = \pm \frac{1}{\sqrt{3}} (J^8_\pm - \sqrt{2} J^{15}_\pm)$ and
$L_\pm = \frac{2}{\sqrt{3}} \left( \mp \frac{1}{2} J^3_\pm
+ \frac{1}{\sqrt{3}} J^8_\pm + \frac{1}{\sqrt{6}} J^{15}_\pm \right)$.

\section{Geometric data on $\mathbb{C}\mathbb{P}^3$}

We present here detailed results of the calculations that are necessary to obtain the main results but are not presented in the main text.

\subsection{Coordinate representation}

The spin connections for the metric \eq{cp3-metric} have the following components
\bea \la{cp3-spin}
&& \omega_{12} = - 2 \cot 2 \rho e^2, \quad \omega_{13} = - \cot \rho e^3, \quad
\omega_{14} = - \cot \rho e^4, \quad \omega_{15} = - \cot \rho e^5, \quad
\omega_{16} = - \cot \rho e^6, \xx
&& \omega_{23} = \cot \rho e^4, \quad \omega_{24} = - \cot \rho e^3, \quad
\omega_{25} = \cot \rho e^6, \quad \omega_{26} = -\cot \rho e^5, \xx
&& \omega_{34} = - \cot \rho e^2 - 2 \csc \rho \cot 2 \theta e^4,
\quad \omega_{35} = - \csc \rho \cot \theta e^5, \quad
\omega_{36} = - \csc \rho \cot \theta e^6, \\
&& \omega_{45} = \csc \rho \cot \theta e^6,
\quad \omega_{46} = - \csc \rho \cot \theta e^5, \xx
&& \omega_{56} = - \cot \rho e^2 - \csc \rho \cot \theta e^4
+ 2 \csc \rho \csc \theta \cot \alpha_1 e^5. \nonumber
\eea
The corresponding curvature tensors can be summarized in a succinct form:
\bea \la{cp3-curv}
&& R_{AB} = 2 P^{ABCD}_- e^C \wedge e^D + 3 I_{AB} \Omega, \xx
&& R_{ABCD} =  (\delta_{AC}\delta_{BD} - \delta_{AD} \delta_{BC})
- \frac{1}{2} \varepsilon^{ABCDEF} I_{EF} + 3 I_{AB} I_{CD},
\eea
where the (almost projection) operator (see Eq. (3.39) in \cite{yang-yun})
\be \la{p+-}
P^{ABCD}_\pm \equiv \frac{1}{4} (\delta_{AC}\delta_{BD}
- \delta_{AD} \delta_{BC}) \pm \frac{1}{8} \varepsilon^{ABCDEF} I_{EF}
\ee
has played an important role in \eq{bogo-action} and \eq{w-hodge}.

We identify $SU(4)$ gauge fields defined by $A^{(+) a} = \omega_{AB} \eta^{(+)a}_{AB}$:
\bea \la{su4-gauge}
&& A^{(+)3} = (\tan \rho - 2 \cot \rho) e^2 - 2 \csc \rho \cot 2 \theta e^4,
\quad A^{(+)6} = - 2 \cot \rho e^6, \quad A^{(+)7} = 2 \cot \rho e^5, \xx
&& A^{(+)8} = - \frac{1}{\sqrt{3}} (\tan \rho + 2 \cot \rho) e^2 +
\frac{1}{\sqrt{3}} \csc \rho (\tan \theta - 3 \cot \theta) e^4 +
\frac{4}{\sqrt{3}} \csc \rho \csc \theta \cot \alpha_1 e^5, \xx
&& A^{(+)11} = - 2 \csc \rho \cot \theta e^6, \quad  A^{(+)12} = 2 \csc \rho \cot \theta e^5,
\quad A^{(+)13} = - 2 \cot \rho e^4, \quad  A^{(+)14} = - 2 \cot \rho e^3, \xx
&& A^{(+)15} = - \frac{4}{\sqrt{6}} \cot 2 \rho e^2 -
\frac{2}{\sqrt{6}} \csc \rho \tan \theta e^4 +
\frac{4}{\sqrt{6}} \csc \rho \csc \theta \cot \alpha_1 e^5,
\eea
and $A^{(+)1} = A^{(+)2} = A^{(+)4} = A^{(+)5} = A^{(+)9} = A^{(+)10} = 0$.
The corresponding field strengths are determined by
$F^{(+) a} = dA^{(+) a} - \frac{1}{2} f^{abc} A^{(+) b} \wedge A^{(+) c}$:
\bea \la{su4-field}
&& F^{(+) 3} = 6 e^1 \wedge e^2 + 6 e^3 \wedge e^4 + 4 e^5 \wedge e^6, \quad
F^{(+) 6} = 2 e^1 \wedge e^6 - 2 e^2 \wedge e^5,  \xx
&& F^{(+) 7} = - 2 e^1 \wedge e^5 - 2 e^2 \wedge e^6,
\quad F^{(+) 8} = \frac{2}{\sqrt{3}} e^1 \wedge e^2
+ \frac{6}{\sqrt{3}} e^3 \wedge e^4 + \frac{8}{\sqrt{3}} e^5 \wedge e^6, \xx
&& F^{(+) 11} = 2 e^3 \wedge e^6 - 2 e^4 \wedge e^5, \quad
F^{(+) 12} = - 2 e^3 \wedge e^5 - 2 e^4 \wedge e^6, \xx
&& F^{(+) 13} = 2 e^1 \wedge e^4 - 2 e^2 \wedge e^3, \quad
F^{(+) 14} =  2 e^1 \wedge e^3 + 2 e^2 \wedge e^4, \xx
&& F^{(+) 15} = \frac{8}{\sqrt{6}} e^1 \wedge e^2
+ \frac{8}{\sqrt{6}} e^5 \wedge e^6,
\eea
and $F^{(+)1} = F^{(+)2} = F^{(+)4} = F^{(+)5} = F^{(+)9} = F^{(+)10} = 0$.

If we instead embed $SU(4)$ gauge fields to $A^{(-) a} = \omega_{AB} \eta^{(-)a}_{AB}$, we get
\bea \la{-su4-gauge}
&& A^{(-)3} = - A^{(+)3}, \quad A^{(-)4} =  A^{(+)11}, \quad
A^{(-)5} = A^{(+)12}, \quad A^{(-)9} = - A^{(+)6}, \xx
&& A^{(-)10} = - A^{(+)7}, \quad A^{(-)13} = - A^{(+)13}, \quad
A^{(-)14} = A^{(+)14}, \xx
&& A^{(-)8} = \frac{1}{\sqrt{3}} (\tan \rho - 2 \cot \rho) e^2 -
\frac{2}{\sqrt{3}} \csc \rho \csc 2 \theta  e^4 +
\frac{4}{\sqrt{3}} \csc \rho \csc \theta \cot \alpha_1 e^5, \xx
&& A^{(-)15} = - \frac{4}{\sqrt{6}} \csc 2 \rho e^2 +
\frac{2}{\sqrt{6}} \csc \rho (\tan \theta - 2 \cot \theta) e^4 +
\frac{4}{\sqrt{6}} \csc \rho \csc \theta \cot \alpha_1 e^5,
\eea
and $A^{(-)1} = A^{(-)2} = A^{(-)6} = A^{(-)7} = A^{(-)11} = A^{(-)12} = 0$.
Their field strengths are
\bea \la{-su4-field}
&& F^{(-)3} = - F^{(+)3}, \quad F^{(-)4} =  F^{(+)11}, \quad
F^{(-)5} = F^{(+)12}, \quad F^{(-)9} = - F^{(+)6}, \xx
&& F^{(-)10} = - F^{(+)7}, \quad F^{(-)13} = - F^{(+)13}, \quad
F^{(-)14} = F^{(+)14}, \xx
&& F^{(-)8} = \frac{6}{\sqrt{3}} e^1 \wedge e^2
+ \frac{2}{\sqrt{3}} e^3 \wedge e^4 + \frac{8}{\sqrt{3}} e^5 \wedge e^6, \xx
&& F^{(-)15} = \frac{8}{\sqrt{6}} e^3 \wedge e^4
+ \frac{8}{\sqrt{6}} e^5 \wedge e^6,
\eea
and $F^{(-)1} = F^{(-)2} = F^{(-)6} = F^{(-)7} = F^{(-)11} = F^{(-)12} = 0$.

\subsection{Coset representation}

We will take the $G$-invariant metric in \eq{coset-metric} as $g_{ij} = \delta_{ij}$ for simplicity
though it can take a more general form \cite{mhs-coset}.

%\noindent
\subsubsection{$\mathbb{C}\mathbb{P}^3 = SU(4)/U(3)$}

The connection one-forms in \eq{co-cps-conn} take the following components
\bea \la{cp3-conn}
&& \omega_{12} = \omega^3 + \frac{1}{\sqrt{3}} \omega^8 + 2 \sqrt{\frac{2}{3}} \omega^{15}, \quad \omega_{13} = \omega^{2}, \quad
\omega_{14} = \omega^{1}, \quad \omega_{15} = \omega^{5}, \quad
\omega_{16} = \omega^{4}, \xx
&& \omega_{23} = - \omega^{1}, \quad \omega_{24} = \omega^{2}, \quad
\omega_{25} = - \omega^{4}, \quad \omega_{26} = \omega^{5}, \xx
&& \omega_{34} = - \omega^3 + \frac{1}{\sqrt{3}} \omega^8 + 2 \sqrt{\frac{2}{3}} \omega^{15},
\quad \omega_{35} = \omega^{7}, \quad \omega_{36} = \omega^{6},
\quad \omega_{45} = - \omega^{6},  \quad \omega_{46} = \omega^{7}, \xx
&& \omega_{56} = \frac{2}{\sqrt{3}} ( - \omega^8 + \sqrt{2} \omega^{15} ).
\eea
The curvature two-forms ${R^i}_j = - \frac{1}{2} {C_{kl}}^{\bar{a}} {C_{\bar{a}j}}^i
\theta^k \wedge \theta^l$ defined by \eq{coset-curv} take exactly the same expression as Eq. \eq{cp3-curv}
if the coset coframes $\theta^i$ are identified with the orthonormal coframes, i.e., $\theta^i \cong e^A, \; i, A = 1, \cdots, 6.$

It is straightforward to identify $SU(4)$ gauge fields defined by $A^{(+) a} = \eta^{(+)a}_{ij} \omega_{ij}, \; a = (\bar{a}, 8+i)$:
\bea \la{cocp3-conn}
&& A^{(+)3} = \frac{2}{\sqrt{3}} (\omega^8 + 2 \sqrt{2} \omega^{15}),
\quad A^{(+)6} = 2 \omega^{4}, \quad A^{(+)7} = - 2 \omega^{5}, \xx
&& A^{(+)8} = - \frac{2}{3} (\sqrt{3} \omega^{3}
+ 2 \omega^{8} - 2 \sqrt{2} \omega^{15} ), \quad
A^{(+)11} = 2 \omega^{6} , \quad  A^{(+)12} = - 2 \omega^{7}, \xx
&& A^{(+)13} = 2 \omega^{1}, \quad  A^{(+)14} = 2 \omega^{2}, \quad
A^{(+)15} = \frac{2 \sqrt{2}}{3} \left( \sqrt{3} \omega^{3}
- \omega^{8} + \sqrt{2} \omega^{15} \right),
\eea
and $A^{(+)1} = A^{(+)2} = A^{(+)4} = A^{(+)5} = A^{(+)9} = A^{(+)10} = 0$.
The corresponding curvature two-forms defined by
$F^{(+) a} = dA^{(+) a} - \frac{1}{2} f^{abc} A^{(+) b} \wedge A^{(+) c}$
also take exactly the same form as \eq{su4-field}
with the identification $\theta^i \cong e^A$.

%\noindent
\subsubsection{Symmetric $\mathbb{C}\mathbb{P}^3 = Sp(4)/[SU(2) \times U(1)]_{max}$}

The connection one-forms in \eq{co-sp4-conn} have the following components
\bea \la{sp4-conn}
&& \omega_{12} = t^{12 \bar{a}} \omega^{\bar{a}} = \omega^{10}, \quad \omega_{13}
= t^{13 \bar{a}} \omega^{\bar{a}} = \omega^9, \quad
\omega_{14} = t^{14 \bar{a}} \omega^{\bar{a}} = 0,
\quad \omega_{15} = t^{15 \bar{a}} \omega^{\bar{a}} = - \omega^8, \xx
&& \omega_{16} = t^{16 \bar{a}} \omega^{\bar{a}} = 0,
\quad \omega_{23} = t^{23 \bar{a}} \omega^{\bar{a}} = 0, \quad
\omega_{24} = t^{24 \bar{a}} \omega^{\bar{a}} = \omega^{9},
\quad \omega_{25} = t^{25 \bar{a}} \omega^{\bar{a}} = 0, \xx
&& \omega_{26} = t^{26 \bar{a}} \omega^{\bar{a}} = - \omega^{8},
\quad \omega_{34} = t^{34 \bar{a}} \omega^{\bar{a}} = \omega^{10},
\quad \omega_{35} = t^{35 \bar{a}} \omega^{\bar{a}} = \omega^{7},
\quad \omega_{36} = t^{36 \bar{a}} \omega^{\bar{a}} = 0, \xx
&& \omega_{45} = t^{45 \bar{a}} \omega^{\bar{a}} = 0,
\quad \omega_{46} = t^{46 \bar{a}} \omega^{\bar{a}} =  \omega^{7},
\quad \omega_{56} = t^{56 \bar{a}} \omega^{\bar{a}} = \omega^{10}.
\eea
The curvature two-forms are defined by \eq{coset-curv} which are compactly written as
\be \la{sp4-curv}
R_{(2\bar{i}-1)(2\bar{j}-1)} = R_{(2\bar{i}) (2\bar{j})}
= \theta^{2\bar{i}-1} \wedge \theta^{2\bar{j}-1}
+ \theta^{2\bar{i}} \wedge \theta^{2 \bar{j}}, \quad
R_{(2\bar{i}-1)(2\bar{j})} = \delta_{\bar{i} \bar{j}} \Omega,
\ee
where $\bar{i}, \bar{j} = 1,2,3$ and
\be \la{sp4-omega}
\Omega = d \omega^{10} = \theta^1 \wedge \theta^2 + \theta^3 \wedge\theta^4 +  \theta^5 \wedge \theta^6 .
\ee

For the reason discussed in the text, we identify $SU(4)$ gauge fields in the anti-chiral sector defined
by $A^{(-)a} = \eta^{(-)a}_{ij} \omega_{ij}, \; a = 1, \cdots, 15$:
\bea \la{cosp4-conn}
&&  A^{(-)3} = -  2 \omega^{10},  \quad A^{(-)4} = 0, \quad  A^{(-)5} = - 2 \omega^{7},
\quad A^{(-)8} = \frac{2}{\sqrt{3}} \omega^{10}, \quad A^{(-)9} = 0, \xx
&&  A^{(-)10} = - 2 \omega^{8}, \quad A^{(-)13} = 0, \quad A^{(-)14} = 2 \omega^{9},
\quad A^{(-)15} = \sqrt{\frac{2}{3}} \omega^{10},
\eea
and $A^{(-)1} = A^{(-)2} = A^{(-)6} = A^{(-)7} = A^{(-)11} = A^{(-)12} = 0$.
Their curvature two-forms are given by
\bea \la{-sp4-curv}
&& F^{(-)3} = - 2 \Omega, \quad F^{(-)4} =  0, \quad
F^{(-)5} = - 2 (\theta^3 \wedge \theta^5 + \theta^4 \wedge \theta^6),
\quad F^{(-)8} = \frac{2}{\sqrt{3}} \Omega,  \xx
&& F^{(-)9} = 0, \quad F^{(-)10} = 2 ( \theta^1 \wedge \theta^5
+ \theta^2 \wedge \theta^6 ), \quad F^{(-)13} = 0, \xx
&& F^{(-)14} = 2 (\theta^1 \wedge \theta^3 + \theta^2 \wedge \theta^4),
\quad F^{(-)15} = \sqrt{\frac{2}{3}} \Omega,
\eea
and $F^{(-)1} = F^{(-)2} = F^{(-)6} = F^{(-)7} = F^{(-)11} = F^{(-)12} = 0$.
Note that the gauge fields $A^{(-)3, 8, 15}$ are basically $U(1)$ gauge fields with field strengths $F^{(-)3, 8, 15} = dA^{(-)3, 8, 15}$.
They determine the $U(1)$ gauge field
defined by \eq{a0} as
\be \la{sp4-u1}
A^{(\pm)0} = 3 \omega^{10}, \qquad F^{(\pm)0} = dA^{(\pm)0} = 3 \Omega.
\ee

Although the coset space $M = Sp(4)/[SU(2) \times U(1)]_{max}$ is not a K\"ahler manifold
but a nearly K\"ahler manifold, $SU(4)$ gauge fields on $M$
behave like $U(3)$ gauge fields. We will see that the non-symmetric homogeneous space
$Sp(4)/[SU(2) \times U(1)]_{non-max}$ reveals a completely different behavior.

%\noindent
\subsubsection{Non-symmetric $\mathbb{C}\mathbb{P}^3 = Sp(4)/[SU(2) \times U(1)]_{non-max}$}

The connection one-forms in \eq{ns-sp4-conn} take the following components
\bea \la{ns-leci-conn}
&& \omega_{12} = - ( \omega^9 + \omega^{10}), \quad \omega_{13}
= - \frac{1}{2} (1 + \alpha) \theta^6, \quad
\omega_{14} = \frac{1}{2} (1 + \alpha) \theta^5,
\quad \omega_{15} = - \omega^8 - \frac{1}{2} (1 + \alpha) \theta^4, \xx
&& \omega_{16} = - \omega^7 + \frac{1}{2} (1 + \alpha) \theta^3,
\quad \omega_{23} = - \frac{1}{2} (1 + \alpha) \theta^5, \quad
\omega_{24} = - \frac{1}{2} (1 + \alpha) \theta^6, \xx
&& \omega_{25} = \omega^{10} + \frac{1}{2} (1 + \alpha) \theta^3,
\quad \omega_{26} = - \omega^{8} + \frac{1}{2} (1 + \alpha) \theta^4,
\quad \omega_{34} = - 2 \omega^{10},
\quad \omega_{35} = - \frac{1}{2} (1 + \alpha) \theta^2, \xx
&& \omega_{36} = - \frac{1}{2} (1 + \alpha) \theta^1,
\quad \omega_{45} = \frac{1}{2} (1 + \alpha) \theta^1,
\quad \omega_{46} = - \frac{1}{2} (1 + \alpha) \theta^6, \quad \omega_{56} = \omega^9,
\eea
where $\alpha$ is the torsion parameter in \eq{mod-tor}.
The curvature two-forms defined by \eq{nc-sp4-curv} take rather complex forms
\bea \la{ns-sp4-curv}
&& R_{ij} = \frac{1}{4} (1-\alpha^2) \theta^i \wedge \theta^j
- \frac{1}{16} (1+\alpha) (3+\alpha) \varepsilon_{ijkl mn} I_{mn} \theta^k \wedge \theta^l, \xx
&& \qquad \mathrm{for} \; [ij] = (13, 14, 23, 24, 35,36, 45, 46), \xx
&& R_{ij} = \left( 1 + \frac{1}{4} (1- \alpha^2) \right) \theta^i \wedge \theta^j
- \frac{1}{2} \left( 1 - \frac{1}{4} (1+\alpha) (3+\alpha) \right)
\varepsilon_{ij 34 kl} \theta^k \wedge \theta^l, \xx
&& \qquad \mathrm{for} \; [ij] = (15, 16, 25, 26), \xx
&& R_{12} = 2 \left( \theta^1 \wedge \theta^2 + \theta^3 \wedge \theta^4 \right)
- \frac{1}{2} (1 + \alpha)^2
\left( \theta^3 \wedge \theta^4 - \theta^5 \wedge \theta^6 \right), \xx
&& R_{34} = 2 \left( \theta^1 \wedge \theta^2 + 2 \theta^3 \wedge \theta^4
+ \theta^5 \wedge \theta^6 \right) - \frac{1}{2} (1 + \alpha)^2
\left( \theta^1 \wedge \theta^2 + \theta^5 \wedge \theta^6 \right), \xx
&& R_{56} = 2 \left( \theta^3 \wedge \theta^4 + \theta^5 \wedge \theta^6 \right)
+ \frac{1}{2} (1 + \alpha)^2
\left( \theta^1 \wedge \theta^2 - \theta^3 \wedge \theta^4 \right).
\eea
The above result reproduces (6.4) in \cite{mhs-coset} when $\alpha = 0$.

The non-symmetric space $M = Sp(4)/[SU(2) \times U(1)]_{non-max}$ is also a nearly
K\"ahler manifold, so we expect that $SU(4)$ gauge fields on $M$
behave differently from the previous cases. It is easy to identify the $SU(4)$ gauge fields on $M$
from the Levi-Civita connections in \eq{ns-sp4-conn} using
$A^{(\pm)a} = \eta^{(\pm)a}_{ij} \omega_{ij}, \; a = 1, \cdots, 15$:
\bea \la{ns-sp4-gauge}
&& A^{(+)1}  = - A^{(-)1} = 0, \quad A^{(+)2} = A^{(-)2} = 0, \quad
A^{(+)3} = - A^{(-)3} = - (\omega^{9} + 3 \omega^{10} ),  \xx
&& A^{(+)4} = A^{(-)11} = 0, \quad A^{(+)5} = A^{(-)12} = 0, \quad
A^{(+)6} = - A^{(-)9} = - 2 \omega^{7}, \xx
&& A^{(+)7} = - A^{(-)10} = 2 \omega^8, \quad
A^{(+)8} = \sqrt{3} (\omega^{9} - \omega^{10}),  \quad
A^{(-)8} = \frac{1}{\sqrt{3}} ( \omega^{9} - \omega^{10}), \xx
&& A^{(+)9} = - A^{(-)6} = - (1 + \alpha) \theta^3,
\quad A^{(+)10} = - A^{(-)7} = - (1 + \alpha) \theta^4, \xx
&& A^{(+)11} = A^{(-)4} = - (1 + \alpha) \theta^1,
\quad A^{(+)12} = A^{(-)5} = (1 + \alpha) \theta^2, \xx
&& A^{(+)13} = - A^{(-)13} = (1 + \alpha) \theta^5, \quad
A^{(+)14} = A^{(-)14} = - (1 + \alpha) \theta^6, \xx
&& A^{(+)15} = 0,     % {\sqrt{\frac{2}{3}}} \omega^{10},
\quad A^{(-)15} = 2 {\sqrt{\frac{2}{3}}} (\omega^{9} - \omega^{10}).
\eea
The $SU(4)$ field strengths defined by
$F^{(\pm) a} = dA^{(\pm) a} - \frac{1}{2} f^{abc} A^{(\pm) b} \wedge A^{(\pm) c}$
precisely coincide with the Riemann curvature tensors in \eq{ns-sp4-curv}
with the identification $F^{(\pm) a} = \eta^{(\pm)a}_{ij} R_{ij}$,
as they should be. We list them with a shortand notation, $[ij] : = \theta^i \wedge \theta^j$:
\bea \la{nssp4-field-comp}
&& F^{(+)1} = - \frac{1}{2} (1+\alpha)^2 ([14] + [23]), \quad
F^{(+)2} = \frac{1}{2} (1+\alpha)^2 ([13] - [24]), \xx
&& F^{(+)3} = 2(2[12] + 3[34] + [56])- \frac{1}{2} (1+\alpha)^2 ([12] + [34])
= - F^{(-)3}, \xx
&& F^{(+)4} = - \frac{1}{2} (1+\alpha)^2 ([36] + [45]), \quad
F^{(+)5} = \frac{1}{2} (1+\alpha)^2 ([35] - [46]), \xx
&& F^{(+)6} = 2([16] - [25]) - \frac{1}{2} (1+\alpha)^2 ([16] - [25]), \xx
&& F^{(+)7} = -2([15] + [26]) + \frac{1}{2} (1+\alpha)^2 ([15] + [26]), \xx
&& F^{(+)8} = 2\sqrt{3} ([34] + [56]) + \frac{1}{2\sqrt{3}} (1+\alpha)^2 ([12] - [34] - 2[56]), \xx
&& F^{(-)8} = \frac{2}{\sqrt{3}} ([34] + [56])
+ \frac{\sqrt{3}}{2} (1+\alpha)^2 \left([12] - [34] + \frac{2}{3} [56] \right), \xx
&& F^{(+)9} = - (1+\alpha) ([16] + [25]), \quad
F^{(+)10} = (1+\alpha) ([15] - [26]), \xx
&& F^{(+)11} = (1+\alpha) ([36] - [45]), \quad
F^{(+)12} = - (1+\alpha) ([35] + [46]), \xx
&& F^{(+)13} = (1+\alpha) ([14] - [23]), \quad
F^{(+)14} = (1+\alpha) ([13] + [24]), \xx
&& F^{(+)15} = \sqrt{\frac{2}{3}} (1+\alpha)^2 ([12] - [34] + [56]), \xx
&& F^{(-)15} = \sqrt{\frac{2}{3}} \left ( 4[34] + 4[56] - (1+\alpha)^2 [56] \right),
\eea
where we did not specifically indicate $(-)$-components which can be easily identified.
Note that the resulting gauge fields have a very different structure from the previous examples.
In particular, they cannot be reduced to $H$-connections; for example, $A^{(+)9, \cdots, 14}$ are not $H$-connections,
except for the case $\alpha = -1$.

\section{$C_3 (E)$}

The third Chern number is defined by \eq{3-chern}.
A curvatue tensor of a $G$-bundle $E$ is a Lie-algebra valued two-form,
i.e. $F \in \Omega^2 (\mathrm{End(E))}$.
Thus it can be decomposed according to the structure \eq{2form-decomp} as
\be \la{decomp-f}
F^{(\pm)} = F^{(\pm)a} T^a_\pm = \sqrt{\frac{2}{3}} F^{(\pm)0} T^0_\pm  + F^{(\pm)\dot{a}} T^{\dot{a}}_\pm + F^{(\pm)\hat{a}} T^{\hat{a}}_\pm,
\ee
where
\bea \la{dec-gen038}
&& T^0_+ = \sqrt{\frac{2}{3}} \left( T^3 + \frac{1}{\sqrt{3}} T^8 + \frac{1}{\sqrt{6}} T^{15} \right) = \frac{i}{2\sqrt{6}} \mathrm{diag} (3,-1,-1,-1), \xx
&& T^0_- = - \sqrt{\frac{2}{3}} \left( - T^3 + \frac{1}{\sqrt{3}} T^8 + \frac{1}{\sqrt{6}} T^{15} \right) = - \frac{i}{2\sqrt{6}} \mathrm{diag} (-1,3,-1,-1), \xx
&& T^{\hat{3}}_\pm = \frac{1}{\sqrt{3}} \left( T^8 - \sqrt{2} T^{15} \right)
= \frac{i}{2} \mathrm{diag} (0,0,-1,1), \xx
&& T^{\hat{8}}_+ = \frac{2}{\sqrt{3}} \left( - \frac{1}{2} T^3 + \frac{1}{\sqrt{3}} T^8 + \frac{1}{\sqrt{6}} T^{15} \right)
= \frac{i}{2\sqrt{3}} \mathrm{diag} (0,2,-1,-1), \xx
&& T^{\hat{8}}_- = - \frac{2}{\sqrt{3}} \left( \frac{1}{2} T^3 + \frac{1}{\sqrt{3}} T^8 + \frac{1}{\sqrt{6}} T^{15} \right)
= - \frac{i}{2\sqrt{3}} \mathrm{diag} (2,0,-1,-1),
\eea
and
\bea \la{lie-gen+-}
&& (T_+^{\hat{1}}, T_+^{\hat{2}}, T_+^{\hat{4}}, T_+^{\hat{5}}, T_+^{\hat{6}}, T_+^{\hat{7}}) = (T^{13}, T^{14}, T^{6}, T^{7}, T^{11}, T^{12}), \xx
&& (T_+^{\dot{1}}, T_+^{\dot{2}}, T_+^{\dot{3}}, T_+^{\dot{4}}, T_+^{\dot{5}}, T_+^{\dot{6}}) = (T^{1}, T^{2}, T^{9}, T^{10}, T^{4}, T^{5}), \xx
&& (T_-^{\hat{1}}, T_-^{\hat{2}}, T_-^{\hat{4}}, T_-^{\hat{5}}, T_-^{\hat{6}}, T_-^{\hat{7}}) = (T^{13}, T^{14}, T^{9}, - T^{10}, - T^{4}, T^{5}), \xx
&& (T_-^{\dot{1}}, T_-^{\dot{2}}, T_-^{\dot{3}}, T_-^{\dot{4}}, T_-^{\dot{5}},
T_-^{\dot{6}}) = (T^{1}, T^{2}, T^{6}, - T^{7}, - T^{11}, T^{12}).
\eea
The $U(3)$ generators $T^{\bar{a}} \; (\bar{a} = 0, 1, \cdots, 8)$
in \eq{dec-gen038} and \eq{lie-gen+-} satisfy the product algebra
\be \la{u3-prod}
T^{\bar{a}}T^{\bar{b}} = - \frac{1}{8} \delta^{\bar{a}\bar{b}} - \frac{1}{2} f^{\bar{a}\bar{b}\bar{c}} T^{\bar{c}} + \frac{i}{2} d^{\bar{a}\bar{b}\bar{c}} T^{\bar{c}}
\ee
with the structure constants
\be \la{st-con-0}
f^{0 \hat{a} \hat{b}} = 0, \quad
d^{000} = \sqrt{\frac{2}{3}}, \quad d^{00\hat{a}} = 0, \quad
d^{0 \hat{a} \hat{b}} = - \frac{1}{\sqrt{6}} \delta^{\hat{a} \hat{b}}.
\ee
The other structure constants $f^{\hat{a} \hat{b} \hat{c}}$ and  $d^{\hat{a}\hat{b}\hat{c}}$ can be read off
from the Tables 1 and 2 in \cite{yang-yun} with the sign convention, $(\hat{1}, \hat{2},\hat{4}, \hat{6}) = (1,2,4,6)$ and $(\hat{3}, \hat{5},\hat{7}, \hat{8}) = -(3,5,7,8)$.
The decompostion \eq{decomp-f} then gives us the expansion of the third Chern class
\bea \la{exp-3chern}
&& \Tr F^{(\pm)} \wedge F^{(\pm)} \wedge F^{(\pm)} \xx
&=& \mp \frac{i}{4} \left( d^{\hat{a} \hat{b} \hat{c}} F^{(\pm)\hat{a}} \wedge F^{(\pm)\hat{b}} \wedge F^{(\pm)\hat{c}}
+ \left(\frac{2}{3} \right)^{\frac{3}{2}} d^{000} F^{(\pm)0} \wedge F^{(\pm)0} \wedge F^{(\pm)0} \right. \xx
&& + d^{\dot{a} \dot{b} \dot{c}} F^{(\pm)\dot{a}} \wedge F^{(\pm)\dot{b}} \wedge F^{(\pm)\dot{c}}
+ 2 d^{00 \hat{a}} F^{(\pm)0} \wedge F^{(\pm) 0} \wedge F^{(\pm)\hat{a}} + 2 d^{00 \dot{a}} F^{(\pm)0} \wedge F^{(\pm) 0} \wedge F^{(\pm)\dot{a}} \xx
&& + \sqrt{6} d^{0 \hat{a} \hat{b}} F^{(\pm)0} \wedge F^{(\pm) \hat{a}} \wedge F^{(\pm)\hat{b}}
+ \sqrt{6} d^{0 \dot{a} \dot{b}} F^{(\pm)0} \wedge F^{(\pm) \dot{a}} \wedge F^{(\pm)\dot{b}} \xx
&& + \sqrt{6} (d^{0 \hat{a} \dot{a}} + d^{0 \dot{a} \hat{a}} ) F^{(\pm)0}  \wedge F^{(\pm)\hat{a}} \wedge F^{(\pm) \dot{a}} \xx
&& \left. + 3 d^{\hat{a} \hat{b} \dot{a}} F^{(\pm)\hat{a}} \wedge F^{(\pm) \hat{b}} \wedge F^{(\pm)\dot{a}}
+ 3 d^{\hat{a} \dot{a} \dot{b}} F^{(\pm) \hat{a}} \wedge F^{(\pm) \dot{a}} \wedge F^{(\pm)\dot{b}} \right),
\eea
where $d^{abc} = \pm 4 i \Tr \left( T^a_\pm T^b_\pm T^c_\pm \right)$ are structure constants of the Lie algebra generators
in \eq{dec-gen038} and \eq{lie-gen+-}. For example, they appear in the
product \eq{u3-prod}. The first part of the expansion in \eq{exp-3chern} describes
the contribution of HYM instantons to the Euler number, and the second part corresponds to the contribution
from $U(1)$ gauge fields. Using the result \eq{exp-3chern},
we can isolate the contribution of the HYM instanton and write the third Chern number as follows:
\bea \la{3chern-dec}
C_3(E) &=& \frac{i}{24 \pi^3} \int_M \Tr F^{(\pm)} \wedge F^{(\pm)} \wedge F^{(\pm)} \xx
&=& \pm \frac{1}{96 \pi^3} \int_M d^{\hat{a} \hat{b} \hat{c}} F^{(\pm)\hat{a}} \wedge F^{(\pm)\hat{b}} \wedge F^{(\pm)\hat{c}} + C^\perp_3(E),
\eea
where $C^\perp_3(E)$ represents the contribution to the Euler number from other parts in the expansion \eq{exp-3chern}
and, in particular, includes the contribution of the $U(1)$ gauge fields.

\end{document}